\documentclass[]{jfm}

\usepackage{graphicx}
\usepackage{subcaption}
\usepackage{newtxtext}
\usepackage{newtxmath}
\usepackage{amssymb}
\usepackage{natbib}
\usepackage{hyperref}
\usepackage{xcolor}
\usepackage[makeroom]{cancel}
\hypersetup{
    colorlinks = true,
    urlcolor   = blue,
    citecolor  = black,
}
\newcommand{\RomanNumeralCaps}[1]
\linenumbers
\newcommand{\varleq}{\mathrel{\vcenter{\offinterlineskip
  \hbox{$<$}\vskip-0.0ex\hbox{\scalebox{1.2}[0.6]{$-$}}}}}

\title{Turbulence-Resolving Integral Simulations for Wall-Bounded Flows}

\author{Tanner Ragan\aff{1} \corresp{\email{ragant@uci.edu}},
  Mark Warnecke\aff{1} \corresp{\email{mwarneck@uci.edu}}, Samuel T. Stout\aff{1} \corresp{\email{stoutst@uci.edu}}
 \and Perry L. Johnson\aff{1} \corresp{\email{perry.johnson@uci.edu}}}

\affiliation{\aff{1}University of California Irvine, Mechanical and Aerospace Engineering Department, Samueli School of Engineering, Irvine, CA, USA}

\begin{document}
\maketitle

\begin{abstract}

The physical fidelity of turbulence models can benefit from a partial resolution of fluctuations, but doing so often comes with an increase in computational cost. To explore this trade-off in the context of wall-bounded flows, this paper introduces a framework for Turbulence-Resolving Integral Simulations (TRIS) with the goal of efficiently resolving the largest motions using a two-dimensional, three component representation of the flow defined by instantaneous wall-normal integrals of velocity and pressure.
Self-sustaining turbulence with qualitatively realistic large-scale structures is demonstrated for TRIS on an open-channel (half-channel) flow configuration using moment-of-momentum integral equations derived from Navier-Stokes with relatively simple closure approximations.
Evidence from Direct Numerical Simulations (DNS) suggests that TRIS can theoretically resolve $35-40\%$ of the turbulent skin friction enhancement for friction Reynolds numbers between $180$ and $5200$, without a noticeable decrease or increase as a function of Reynolds number.
The current implementation of TRIS can match this resolution while simulating one flow through time in $\sim$1 minute on a single processor, even for very large Reynolds numbers.
The framework facilitates a detailed apples-to-apples comparison of predicted statistics against data from DNS. Comparisons at friction Reynolds numbers of $395$ and $590$ show that TRIS generates a relatively accurate representation of the flow, while highlight discrepancies that demonstrate a need for improving the closure models. The present results for open-channel flow represent a proof of concept for TRIS as a new approach for wall-bounded turbulence modeling, motivating extension to more general flow configurations such as boundary layers on immersed objects.


\end{abstract}


\section{Introduction}
\label{sec:introduction}
The wide range of scales involved in turbulent boundary layers and other forms of wall-bounded turbulence, common to many engineering and natural flows, presents a difficult challenge to computational modeling and prediction efforts.
The cost of Direct Numerical Simulations (DNS) rises rapidly with increasing Reynolds number, making its use for practical applications computationally infeasible for the foreseeable future.
The Large-Eddy Simulation (LES) framework, meanwhile, provides a potential alternative, but wall-resolved LES (WR-LES) remains quite costly at high Reynolds numbers \citep{Spalart2000, Choi2012, Yang2021}.
Consequently, Reynolds-Averaged Navier-Stokes (RANS) models remain relevant and popular. 
Even with wall-modeled LES (WM-LES) or hybrid RANS-LES techniques, scale-resolving simulations can be costly (even prohibitively so) for turbulent boundary layers and immersed bodies at large Reynolds number \citep{Goc2020}.

RANS-based integral methods for turbulent boundary layers, which pre-date the explosion of computer performance over the past half-century \citep{Kline1968}, provide a significant reduction in computing cost by seeking a solution for (averaged) quantities integrated in the wall-normal direction across the turbulent boundary layer.
For aerodynamic and hydrodynamic boundary layers over immersed bodies, integral methods can be coupled with potential flow solvers to provide rapid prediction, albeit at reduced physical fidelity, e.g., \cite{Drela1989}.
The use of depth-averaged equations is similarly common in many other wall-bounded turbulence scenarios, e.g., \citet{Ungarish2009}.

Compared to RANS-based methods, approaches that partially resolve turbulent fluctuations (e.g., LES) potentially offer a substantial advantage in physical fidelity because of their inherent ability to capture nonlocal behavior in the large-scale motions. 
Large-scale motions (LSMs) and very-large-scale motions (VLSMs), sometimes referred to as superstructures, play a prominent role in wall-bounded turbulence. 
Compared to the turbulent layer thickness (i.e., boundary layer thickness, channel height, pipe radius, etc.), the streamwise lengths of these motions are comparable to and much longer and their wall-normal heights are on the order of the turbulent layer thickness \citep{Brown1977,Monty2007,Monty2009,Lee2017}. 
Furthermore, these large streaky structures possess a significant portion of both the Reynolds shear stress and turbulent kinetic energy \citep{Guala2006,Balakumar2007}. 
The prominence and significance of these structures raises a motivation for numerical approaches to develop a framework around (V)LSMs. 
To successfully develop a reduced-order model, it is imperative to encapsulate the essential dynamics of these structures.

The dynamics of streamwise-oriented fluctuations in wall-bounded turbulence received more earlier attention in the context of buffer-layer streaks in the near-wall region \citep{Kline1967}. 
It was shown, using DNS on a minimal flow unit, that these turbulent motions of the low-speed structures self-sustain independent of the flow in the outer region \citep{Jimenez1991, Jimenez1999}. 
Further DNS analysis demonstrated the presence of streamwise counter-rotating vortices \citep{Kim1987}. 
The interplay between streamwise rolls and streaks is crucial for the self-sustaining process of near-wall streaks \citep{Jimenez1991}. 
Specifically, these rolls produce a ``lift-up'' effect that generates the observed low/high-speed streaks. Due to instabilities and/or nonlinear interactions, the streaks become wavy and breakdown. 
Then, by nonlinear physical mechanisms that result in the breakdown of the streaks, new streamwise vortices are formed to repeat the cycle through another iteration. 
This is the classical understanding of a self-sustaining process for buffer-layer streaks in the near-wall region \citep{Hamilton_Kim_Waleffe_1995}.

Adding the log-layer and outer region to consideration, the role of the (V)LSMs introduces additional complexities. 
The fundamental mechanism for producing and sustaining large-scale turbulence has been a topic of significant interest.
One possibility is that the (V)LSMs require the interaction with the flow in the inner region. \citet{Kim1999} hypothesized that the compilation of hairpin vortices merge to develop the large-scale structures.
Studies using particle image velocimetry have made some observations to this effect \citep{Adrian2000, Jodai_Elsinga_2016}, and
other DNS studies have also observed merging of the small-scale structures to form larger ones \citep{Toh2005,Hwang_Lee_Sung_Zaki_2016}.
However, significant evidence is now available suggesting that there exists a similar self-sustaining mechanism for large and very-large scale motions \citep{Cossu2017, Lee_Moser_2019, Zhou_Xu_Jiménez_2022}. For example, \cite{Hwang_Bengana_2016} observed that the dynamical structure of the large-scale streaks are strikingly similar to that of the near-wall streaks.
Even though a wealth of knowledge on (V)LSMs is provided by these and other studies, encapsulating these motions in a reduced-order modeling framework remains a significant opportunity.

This paper explores the possibility of Turbulence-Resolving Integral Simulations (TRIS), that is, LES-like integral methods in which the wall-normal integration is instantaneously carried out across the entire turbulent layer thickness.
By reducing the description of the turbulent flow from three to two dimensions, a significant cost savings may be possible while still capturing the physics of (V)LSMs.
To the authors' knowledge, this instantaneous wall-normal integral approach has not been previously proposed or investigated, as it differs substantially from both LES and the common RANS-based integral methods. 
At the outset, however, it is not clear (i) how much of the turbulence can still be captured in this two-dimensional (2D) representation, and (ii) how unsteady, 2D, Navier-Stokes-based evolution equations can be developed which support self-sustaining turbulence with realistic structure.
This paper provides an answer for these two questions in the simplified context of an open-channel (half-channel) flow. 

The rest of the paper is organized as follows. The governing equations for TRIS are introduced for an open-channel flow configuration in Section \ref{sec:equation_formulation}. Then, Section \ref{sec:AMI} estimates the theoretical ability of TRIS to capture a certain fraction of active turbulence using open-channel and full-channel DNS data with $180 \leq Re_\tau \leq 5200$. Closure approximations for the TRIS equations are introduced in Section \ref{sec:closure_approximation}. Results from their implementation in Section \ref{sec:results} demonstrate self-sustaining fluctuations with realistic structure, allowing a detailed apples-to-apples evaluation against open-channel DNS data for $Re_\tau = 395$ and $590$. A concluding discussion is provided in Section \ref{sec:conclusion}.

\section{Instantaneous Moment-of-Momentum Evolution Equations}
\label{sec:equation_formulation}

In this section, evolution equations for TRIS are derived from the Navier-Stokes equations and boundary conditions for an open-channel flow,
a useful surrogate with a similar structure to boundary layers.
This configuration shares the general characteristics of wall-bounded turbulence while allowing for homogeneity in the wall-parallel direction and simple boundary conditions at the top of the domain (opposite to the no-slip wall). This choice provides a simple starting point for developing the TRIS framework without the added complexities of spatial development and interaction with a potential flow; the details for including these effects in the context of external boundary layer flows are an important topic for future work.

To guide the derivation, the notation will distinguish between wall-normal and wall-parallel components.
Using a 2D index notation, the indices correspond to the streamwise ($i=1$) and spanwise ($i=2$) directions, such that implied summation only applies over the wall-parallel directions. Therefore, the notation used here is $u_1=u$, $u_2=w$, and $v$ along with $x_1=x$, $x_2=z$, and $y$ for the streamwise, spanwise, and wall-normal components of velocity and position, respectively.
The conservation equations for incompressible flow are non-dimensionalized by the height of the open-channel ($h$) and friction velocity ($u_\tau$).
The bottom wall of the open-channel ($y=0$) is a no-slip, no-penetration boundary whereas a no-penetration, zero-vorticity boundary condition is applied at the top wall ($y=1$).
Using this notation, the incompressible Navier-Stokes equations for conservation of mass, wall-parallel momentum and wall-normal momentum are,
\begin{equation}
    \frac{\partial {u_i}}{\partial x_i} 
    + \frac{\partial {v}}{\partial y} 
    = 0,
    \label{eq:mass}
\end{equation}
\begin{equation}
    \frac{\partial {u_i}}{\partial t} 
    + \frac{\partial {u_i u_j}}{\partial x_j} 
    + \frac{\partial {u_i v}}{\partial y} 
    = -\frac{\partial {p}}{\partial x_i}
    +\frac{1}{Re_{\tau}}\left( \frac{\partial^2 {u_i}}{\partial x_j \partial x_j} + \frac{\partial^2 {u_i}}{\partial y^2} \right)
    + \delta_{i1},
    \label{eq:xzmom}
\end{equation}
\begin{equation}
    \frac{\partial {v}}{\partial t}
    + \frac{\partial {v u_j}}{\partial x_j} 
    + \frac{\partial {v v}}{\partial y}
    = -\frac{\partial {p}}{\partial y}
    + \frac{1}{Re_{\tau}} \left( \frac{\partial^2 {v}}{\partial x_j \partial x_j} 
    + \frac{\partial^2 {v}}{\partial y^2} \right).
    \label{eq:ymom}
\end{equation}
Here, $Re_\tau = \rho u_\tau h / \mu$ is the friction Reynolds number for a fluid with density $\rho$ and viscosity $\mu$. The Kronecker delta, $\delta_{i1}$, is the dimensionless imposed pressure gradient that drives the flow and $p$ is the dimensionless pressure field that enforces \eqref{eq:mass}.

Flow fields are integrated in the wall-normal direction and represented as zeroth and first moments, respectively,
\begin{equation}
    \left\langle \phi \right\rangle_0(x_1, x_2, t) 
    = \int_0^1 \phi(x_1, y, x_2, t) dy,
    \hspace{0.03\linewidth}
    \left\langle \phi \right\rangle_1(x_1, x_2, t)
    = \int_0^1 2y\phi(x_1,y,x_2,t) dy.
    \label{eq:moments-def}
\end{equation}
Thus, the zeroth moment is an unweighted wall-normal integral and the first moment is a wall-normal integral linearly weighted by wall-normal distance to favor events occurring further from the wall. 
In addition to wall-normal integration, the fields are also low-pass filtered in the wall-parallel directions, $\widetilde{\phi}$, where the filter width corresponds to the 2D grid spacing to be used for TRIS.

The filtered zeroth and first moments of \eqref{eq:mass} are
\begin{equation}
    \frac{\partial \langle \widetilde{u}_i \rangle_0}{\partial x_i} 
    = 0,
    \hspace{0.2\linewidth}
    \frac{\partial \langle \widetilde{u}_i \rangle_1}{\partial x_i} 
    = 2 \langle \widetilde{v} \rangle_0,
    \label{eq:moments-of-mass}
\end{equation}
where the zeroth moment of the wall-parallel velocity, $\langle \widetilde{u}_i \rangle_0$, is a two-dimensional, two-component (2D/2C) vector field with zero divergence. 
The no-penetration condition at the upper boundary has been applied in the first moment of mass equation.
The first moment of the wall-parallel velocity, $\langle \widetilde{u}_i \rangle_1$, is also 2D/2C and has divergence equal to (twice) the zeroth moment of the wall-normal velocity, $\langle \widetilde{v} \rangle_0$.
In this way, the inclusion of the first moment provides a three-component (2D/3C) description of the zeroth-moment velocity field.

\begin{figure}
    \centerline{\includegraphics[trim={0 1mm 0 2mm},clip,width=1.0\linewidth]{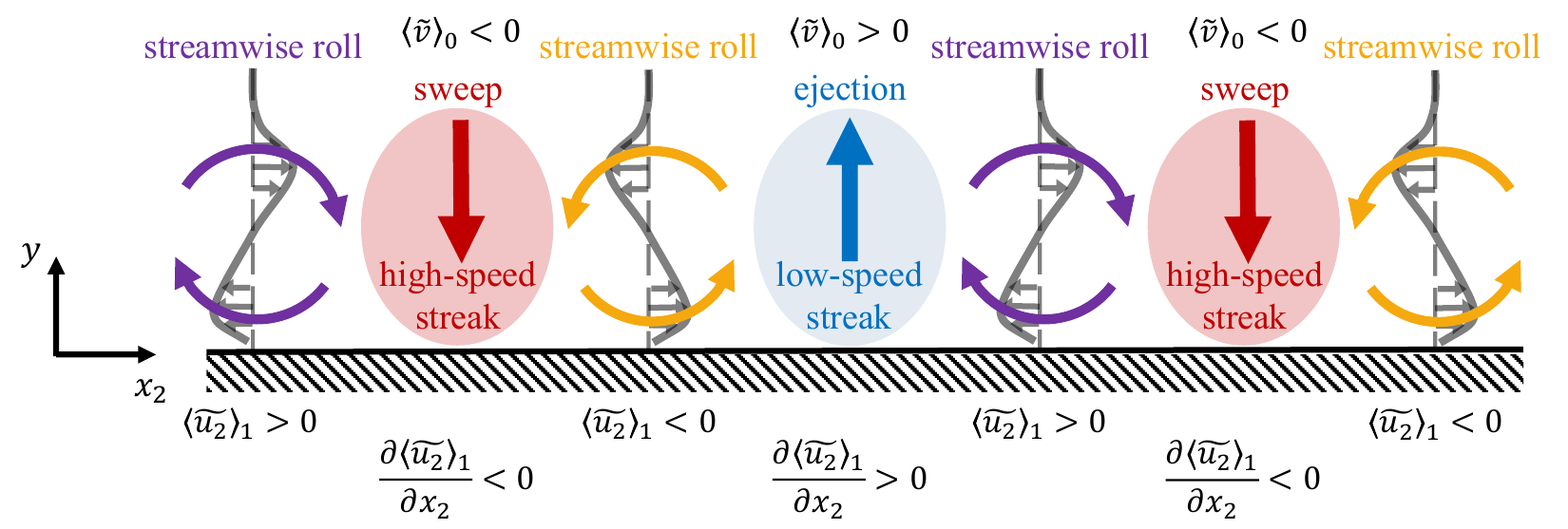}}
    \captionsetup{justification=raggedright, singlelinecheck=false,width=\textwidth}
    \caption{View in the flow direction of the large-scale streamwise rolls generating regions of high and low speed streaks, which correspond to sweeps and ejections, respectively.  The profiles located at the streamwise rolls represent the local spanwise velocity. This phenomena encapsulates the effect of \eqref{eq:moments-of-mass} (right).}
    \label{fig:LSM_rolls}
\end{figure}

The first moment of mass conservation, \eqref{eq:moments-of-mass}, compactly describes the relationship between streamwise rolls and sweeps/ejections (characterized by $\partial \langle \widetilde{u}_2 \rangle_1 / \partial x_2$ and $\left< \widetilde{v} \right>_0$, respectively) that are responsible for the formation of streamwise-oriented streaks. 
This relationship is illustrated in figure\ \ref{fig:LSM_rolls}, which views a streamwise roll-streak flow pattern in the spanwise-normal plane. 
In this view, a clockwise roll corresponds to a positive first-moment of spanwise velocity, $\langle \widetilde{u}_2 \rangle_1 > 0$, and a counter-clockwise roll corresponds to $\langle \widetilde{u}_2 \rangle_1 < 0$. 
The negative or positive gradients in between two counter-rotating rolls correspond to negative or positive wall-normal velocities, in accordance with \eqref{eq:moments-of-mass}.
These sweeps and ejections, respectively, transport fluid across the mean velocity gradient leading to high and low speed streaks in between the rolls.
Thus, the implementation of \eqref{eq:moments-of-mass} allows for the proper relationship between streamwise rolls and streaks that participate in the classical picture of the self-sustaining process.


The dynamics of the zeroth moment velocity fields is given by the zeroth moment of \eqref{eq:xzmom}, also including the wall-parallel filtering,
\begin{equation}
    \frac{\partial \left< \widetilde{u}_i \right>_0}{\partial t} 
	+ \frac{\partial \left< \widetilde{u_i u_j} \right>_0}{\partial x_j}
	= - \frac{\partial \left< \widetilde{p} \right>_0}{\partial x_i}
	+ \frac{1}{Re_\tau} \frac{\partial^2 \left< \widetilde{u}_i \right>_0}{\partial x_j \partial x_j}
	- \widetilde{\tau}_i 
    +\delta_{i1},
    \label{eq:zeroth-moment-momentum}
\end{equation}
where $\widetilde{\tau}_i$ is the instantaneous (dimensionless) wall shear stress (2D/2C). Equation \eqref{eq:zeroth-moment-momentum} lacks an explicit representation of the turbulent momentum flux in the wall-normal direction. 
The dynamics of the first moment velocity field is obtained by applying the first moment to \eqref{eq:xzmom} along with wall-parallel filtering,
\begin{equation}
    \frac{\partial \left< \widetilde{u_i} \right>_1}{\partial t} 
	+ \frac{\partial \left< \widetilde{u_i u_j} \right>_1}{\partial x_j}
	= 2\left< \widetilde{u_i v} \right>_0
	- \frac{\partial \left< \widetilde{p} \right>_1}{\partial x_i}
	+ \frac{1}{Re_\tau} \frac{\partial^2 \left< \widetilde{u_i} \right>_1}{\partial x_j \partial x_j}
	- \frac{2 \widetilde{u}_{i,\text{top}}}{Re_\tau}
	+ \delta_{i1},
    \label{eq:first-moment-momentum}
\end{equation}
where $\widetilde{u}_{i,\text{top}}$ is the wall-parallel velocity vector (2D/2C) at the top of the open-channel (where a Neumann BC is imposed on wall-parallel velocity components and a no-penetration condition is imposed on the wall-normal component).
The first moment of momentum equation includes an explicit representation of turbulent wall-normal momentum flux, namely, the zeroth moment of the Reynolds shear stress, $\langle \widetilde{u_iv} \rangle_0$.

Equations \eqref{eq:zeroth-moment-momentum} and \eqref{eq:first-moment-momentum} contain unclosed terms that need to be modeled: $\widetilde{\tau}_i$, $\left< \widetilde{u_i u_j}\right>_0$, $\left< \widetilde{u_i u_j}\right>_1$, $\left< \widetilde{u_i v}\right>_0$, and $\widetilde{u}_{i,\text{top}}$. 
The divergence of \eqref{eq:zeroth-moment-momentum} and \eqref{eq:first-moment-momentum} provide elliptic equations for the pressure moments, $\langle \widetilde{p} \rangle_0$ and $\langle \widetilde{p} \rangle_1$,
\begin{equation}
    \frac{\partial^2 \langle \widetilde{p} \rangle_0}{\partial x_j \partial x_j} = - \frac{\partial^2 \langle \widetilde{u_i u_j} \rangle_0}{\partial x_i \partial x_j} - \frac{\partial \widetilde{\tau}_j}{\partial x_j},
    \label{eq:pressure_poisson_zero_mom}
\end{equation}
\begin{equation}
    \frac{\partial^2 \langle \widetilde{p} \rangle_1}{\partial x_j \partial x_j} = 2[\widetilde{p}_{\text{top}} -\widetilde{p}_{\text{bot}}] - \frac{\partial^2 \langle \widetilde{u_i u_j} \rangle_1}{\partial x_i \partial x_j} + 4 \frac{\partial \langle \widetilde{u_i v} \rangle_0}{\partial x_i}.
    \label{eq:pressure_poisson_first_mom}
\end{equation}
The zeroth moment of \eqref{eq:ymom} has been used to simplify \eqref{eq:pressure_poisson_first_mom}.
Here, $\widetilde{p}_{\text{top}}$ and $\widetilde{p}_{\text{bot}}$ are the pressures at the top and bottom of the open-channel, respectively, and their difference must be modeled to close the system of equations. 

\section{Turbulence Resolution Estimate}
\label{sec:AMI}

Before introducing closure models for \eqref{eq:zeroth-moment-momentum}-\eqref{eq:pressure_poisson_first_mom}, it is worthwhile to consider how much of the turbulent motions can theoretically be captured in this 2D/3C representation of the flow. This is accomplished by asking a more specific question in terms of the turbulent enhancement of the skin friction coefficient, $C_f = 2 / \overline{u}_{\text{top}}^2$, relative to a laminar flow with the same $Re_{\text{top}} = \overline{u}_{\text{top}} Re_\tau$. The $\overline{\phi}$ operator denotes a Reynolds average, and the fluctuation about the mean is $\phi^{\prime} = \phi - \overline{\phi}$.

Subtracting \eqref{eq:first-moment-momentum} from \eqref{eq:zeroth-moment-momentum} and averaging, assuming statistical stationarity and homogeneity in the wall-parallel directions,
\begin{equation}
    1 = 2\frac{\overline{u}_{\text{top}}}{Re_\tau} + 2\left<-\overline{u^\prime v^\prime}\right>_0
    = \frac{\frac{4}{Re_{\text{top}}} + \frac{\left<-4\overline{u^\prime v^\prime}\right>_0}{\overline{u}_{\text{top}}^2}}{C_f}
    = \frac{C_{f,\text{lam}} + C_{f,\text{turb}}}{C_f},
    \label{eq:AMI}
\end{equation}
where $C_{f,\text{lam}} = 4 Re_{\text{top}}^{-1}$ is the skin friction coefficient of a laminar open-channel flow, therefore $C_{f,\text{turb}} = - 4 \left<\overline{u^\prime v^\prime}\right>_0 / \overline{u}_{\text{top}}^2$ represents the turbulent enhancement of the skin friction coefficient relative to the laminar state. 
This type of equation was previously introduced as the Angular Momentum Integral (AMI) equation by \citet{Elnahhas_2022} for spatially-developing boundary layers. 
The turbulent enhancement, $C_{f,\text{turb}}$, is partially resolved by the 2D/3C zeroth moment velocity vector field, $\left<\overline{u^\prime v^\prime}\right>_0 = \overline{\left< \widetilde{u} \right>_0^\prime \left< \widetilde{v} \right>_0^\prime} + \left<\overline{u^{\prime\prime} v^{\prime\prime}}\right>_0$, where $\phi^{\prime\prime} = \phi^\prime - \langle \widetilde{\phi} \rangle_0^\prime$ is the unresolved portion of a fluctuating field when integrated in the wall-normal direction and filtered in the wall-parallel directions. Thus, the turbulent skin friction enhancement is the sum of a resolved and unresolved portion, $C_{f,\text{turb}} = C_{f,\text{res}} + C_{f,\text{unres}}$. 

Note that the skin friction in terms of the velocity at the top of the open-channel domain corresponds more closely to the typical definition for boundary layer flows, whereas friction factors based on the bulk velocity (flow rate) are more common for internal flows such as pipes and channels. The reason for the choice depends on the context. Internal flows are typically analyzed in terms of flow rates whereas the potential flow velocity (or edge velocity) is more relevant for boundary layer contexts. Thus, the choice of this $C_f$ definition reflects an interest in analyzing the open-channel flow as one would treat the engineering context of an external boundary layer.
The alternative choice to analyze the open-channel flow in terms of friction factor based on bulk velocity (flow rate) would lead to the use of the Fukagata-Iwamoto-Kasagi (FIK) equation \citep{Fukagata2002, Nikora2019, Duan2021}, which is derived from the second moment of momentum equation and contains the first moment of the Reynolds shear stress. \citet{Elnahhas_2022} discuss the similarities and differences between the AMI and FIK equations in more detail.

\begin{figure}
    \centerline{\includegraphics{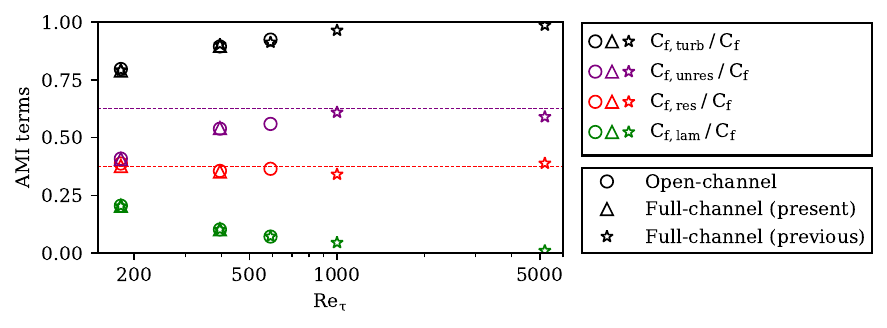}}
    \captionsetup{justification=raggedright, singlelinecheck=false,width=\textwidth}
    \caption{Decomposition of the total skin friction in \eqref{eq:AMI}. The circular and triangular markers associate with the authors' open-channel and full-channel flow simulations, respectively. The star marker corresponds to full-channel flow simulations from previous work \citep{Moser1999, Lee2015, Graham2016}. The colors black, purple, red, and green represents the total, unresolved, and resolved skin friction by turbulent enhancement, and laminar skin friction, respectively. The purple and red dashed lines are at values of $\frac{5}{8}$ and $\frac{3}{8}$, respectively.}
    \label{fig:AMI_balance}
\end{figure}

DNS was used to compute the terms in the AMI equation, including the resolved and unresolved components of the Reynolds shear stress, for $180 \varleq Re_\tau \varleq 5200$ on a domain size of $L_x=8\pi$ and $L_z=3\pi$. The lower friction Reynolds number flows ($Re_\tau = 180, 395, 590$) were simulated using a second-order, staggered finite difference code \citep{Lozano2018} for both full-channel and open-channel configurations and the data for the larger Reynolds numbers ($Re_\tau = 1000, 5200$) are full-channel simulations from the Johns Hopkins Turbulence Databases \citep{Graham2016, Lee2015}. Note that \eqref{eq:AMI} is equally valid for the bottom half of a full-channel flow where $\overline{u}_{\text{top}}$ corresponds to the average centerline velocity and $h$ is the half-height of the full-channel. The results using a spectral cutoff filter with $k_\text{cut} h=16$ applied to the streamwise and spanwise directions are shown in figure\ \ref{fig:AMI_balance}. In this analysis, the isotropic cutoff filter of $k_\text{cut}h=16$ is chosen based on the grid resolution used in Section \ref{sec:results}. Finer wall-parallel resolution does not significantly increase $C_{f,\text{res}}$ \citep{Ragan2025}. The estimated statistical convergence error, following \citet{Shirian2023}, is approximately equal to or smaller than the size of the symbols used, and is further discussed in Appendix \ref{sec:DNSverification}. There is no noticeable difference between open-channel and full-channel flow results for friction Reynolds numbers $180$ to $590$, providing the basis to use full-channel flow datasets at higher Reynolds numbers ($1000$ and $5200$) to estimate the AMI results for the open-channel configuration.

As expected, the laminar skin friction decays with increasing $Re_\tau$, so the sum of the resolved and unresolved turbulent enhancement approaches unity. The unresolved portion generally increases with $Re_\tau$ and appears to approach an asymptotic value based on the available data. Assuming the observed trends continue, it may be estimated that the TRIS equations derived in \S\ref{sec:equation_formulation} can resolve approximately $35-40\%$ of the turbulent skin friction enhancement at large Reynolds numbers using a numerical resolution of $\sim h/5$ in the wall-parallel directions.
This theoretical estimate is based on DNS data for open-channel and full-channel flows, and it is not tied to any particular implementation of closure models for TRIS.


\section{Closure Approximations}
\label{sec:closure_approximation}

To perform a TRIS calculation, a few unclosed terms in \eqref{eq:zeroth-moment-momentum}-\eqref{eq:pressure_poisson_first_mom} need to be approximated in terms of the resolved 2D fields.
The present objective is to demonstrate the sufficiency of the framework presented in \S\ref{sec:equation_formulation} for supporting self-sustaining turbulent fluctuations with realistic structure.
To this end, a simple closure is presented by writing the wall-parallel velocity as $\widetilde{u}_i = U_i + U_i^{\prime\prime}$, where $U_i(x_1, y, x_2, t)$ is interpreted as the mean velocity profile conditioned on the local resolved state defined by $\langle \widetilde{u_i} \rangle_0(x_1, x_2, t)$ and $\langle \widetilde{u_i} \rangle_1(x_1, x_2, t)$. 
For now, this conditional mean profile is modeled using a skewed Coles profile \citep{Coles_1956}, 
\begin{equation}
    U_i = \left[ \frac{1}{\kappa} \text{ln} y + \left( \frac{1}{\kappa} \text{ln}Re_{\ast} + B \right) \right] e_{i,\ast} + \left[ \frac{2 \Pi}{\kappa} \text{sin}^2\left(\frac{\pi}{2}y\right)  \right] e_{i,\Pi},
    \label{eq:coles_profile}
\end{equation}
where $\kappa=0.41$ is the inverse log slope and $B$ is the log vertical intercept, parameters that are pre-set. The local friction Reynolds number is $Re_{\ast}(x_1, x_2, t)$ and $\Pi(x_1, x_2, t)$ is the local wake parameter. The 2D unit vectors $e_{i,\ast}(x_1, x_2, t)$ and $e_{i,\Pi}(x_1, x_2, t)$ align with the local wall shear stress and wake correction, respectively, allowing for skew in the instantaneous velocity profile. Specific details on the alignments is provided in Appendix \ref{sec:derive_model}.
The local values of $Re_*$, $\Pi$, $e_{i,\ast}$, and $e_{i,\Pi}$ are uniquely determined at each point in the $x_1-x_2$ plane given the local resolved state defined by the zeroth and first moments, $\left< \widetilde{u}_i \right>_0 = \left< U_i \right>_0$ and $\left< \widetilde{u}_i \right>_1 = \left< U_i \right>_1$, where it is assumed that the integral moments of the fluctuations about the conditionally averaged velocity profiles are neglected.


Equation \eqref{eq:coles_profile} is evaluated at $y=1$ to close $\widetilde{u}_{i,\text{top}}$. The local wall shear stress $\widetilde{\tau_i}$ is tied to the log portion of the assumed profile in \eqref{eq:coles_profile},
\begin{equation}
    \widetilde{\tau_i} 
    = \left(\frac{Re_\ast}{Re_\tau}\right)^2 e_{i,\ast}.
    \label{tau_i}
\end{equation}
The zeroth and first moments of $\widetilde{u_i u_j}$ are closed by decomposing the term into a portion that is resolved by the Coles profile and unresolved ($\sigma_{0,ij}$ and $\sigma_{1,ij}$), 
\begin{equation}
    \langle \widetilde{u_i u_j} \rangle_0 = \langle U_i U_j \rangle_0 + \sigma_{0,ij},
	\hspace{0.2\linewidth} 
    \langle \widetilde{u_i u_j} \rangle_1 = \langle U_i U_j \rangle_1 + \sigma_{1,ij}.
    \label{eq:wall-parallel_nonlinear_decomposition}
\end{equation}
The resolved portion can be directly computed from the assumed profile, while the unresolved part is modeled by a wall-parallel eddy viscosity approximation \citep{Smag1963}, $\sigma_{0,ij} = C_s \Delta^2 \sqrt{\langle S_{mn} \rangle_{0} \langle S_{mn} \rangle_0} \langle S_{ij} \rangle_{0}$ and $\sigma_{1,ij} = C_s \Delta^2 \sqrt{\langle S_{mn} \rangle_{1} \langle S_{mn} \rangle_1} \langle S_{ij} \rangle_{1}$, where $C_s = 0.78$ is chosen to be a small value that is still large enough to ensure stability, $S_{ij}$ is the wall-parallel strain rate tensor, and $\Delta$ is the grid spacing (filter width).
The resolved portions, $\langle U_i U_j \rangle_0$ and $\langle U_i U_j \rangle_1$, are linearized about reference (mean) values for the log offset, $B$, and the Coles wake strength, $\Pi_{\text{ref}}$, which are set to match the mean zeroth and first moment of the streamwise velocity computed from DNS.

The $\left< \widetilde{u_i v} \right>_0$ term is also decomposed into a resolved and unresolved component,
\begin{equation}
    \left< \widetilde{u_i v} \right>_0 
    = \langle U_i \rangle_0 \langle \widetilde{v} \rangle_0 
    + \langle \widetilde{u''_i v''} \rangle_0.
    \label{eq:reynolds_shear_decomposition}
\end{equation}
Here, {$\langle \widetilde{u''_i v''} \rangle_0$} is modeled using an eddy viscosity with an attached eddy scaling,
\begin{equation}
    -2\langle \widetilde{u''_i v''} \rangle_0 = C_{uv} \frac{\left[1+(1-C_{\Pi})\Pi_{\text{ref}}\right]e_{i,\ast} + C_{\Pi}\Pi e_{i,\Pi}}{1 + \Pi_{\text{ref}}},
\end{equation}
where $C_{uv}$ is set using the AMI balance (the purple symbols in figure\ \ref{fig:AMI_balance}) and $C_\Pi$ is tuned to allow for the correct amount of resolved Reynolds shear stress, $\langle U_i \rangle_0 \langle \widetilde{v} \rangle_0$ (the red symbols in figure\ \ref{fig:AMI_balance}).
Lastly, the pressure difference between the top and bottom of the open-channel is closed by assuming a linear pressure profile at each location, {$\widetilde{p}_{\text{top}} - \widetilde{p}_{\text{bot}} = 6\left[\langle \widetilde{p} \rangle_1 - \langle \widetilde{p} \rangle_0\right]$}. A detailed derivation of these closure approximations is available in Appendix \ref{sec:derive_model}.

\section{Results}
\label{sec:results}

\subsection{Numerical Implementation and Self-Sustaining Turbulence}
A Python code was developed to solve \eqref{eq:zeroth-moment-momentum}-\eqref{eq:pressure_poisson_first_mom} together with the closure models described in \S\ref{sec:closure_approximation} using a pseudo-spectral approach on a doubly-periodic domain of size $L_x = 8\pi$ and $L_z = 3\pi$ to match the DNS domain. The maximum dimensionless wavenumber is $k_\text{max} h = k_\text{cut} h = 16$ in the wall-parallel directions, such that the grid spacing based on collocation points is $\Delta = \pi/16 \approx \frac{1}{5}$ (five grid points per half-channel thickness). Initially ($t=0$), $\langle \widetilde{u}\rangle_0$ is set to a uniform field based on the approximate mean velocity, $\langle \widetilde{u}\rangle_1$ and $\langle \widetilde{w}\rangle_0$ are initialized to zero, while $\langle \widetilde{w}\rangle_1$ is initialized with white noise.
As the simulation advances from the structureless initial conditions, a statistically stationary state emerges with self-sustaining fluctuations. The details of the structure and statistics observed in TRIS simulations are shown below. For now, it is emphasized that the TRIS equations comprised of instantaneous zeroth and first order moments of momentum described in \S\ref{sec:equation_formulation} produce self-sustaining fluctuations when used with the relatively straightforward closures described in \S\ref{sec:closure_approximation}. Earlier attempts by the authors to generate self-sustaining turbulence without the first moment equation were unsuccessful.

\subsection{Choice and Verification of Model Parameters}

Results in this section are presented for a choice of model parameters shown in table \ref{tab:tune_params}. 
These model parameters are tuned to provide accurate values for the results reported in table \ref{tab:stat_verification}. Note that the parameter tuning at $Re_\tau = 1000$ and $5200$ relies on full-channel DNS. Results in figure\ \ref{fig:AMI_balance} verify that full-channel DNS data provides a good proxy for open-channel flow for the quantities used in the parameter tuning.

\begin{table}
    \captionsetup{justification=raggedright, singlelinecheck=false,width=\textwidth}
    \caption{Values of tuning (and set) parameters in TRIS at various $Re_\tau$ (established in \S\ref{sec:closure_approximation}, additional details provided in Appendix \ref{sec:derive_model}).}
    \vspace{2mm}
    \begin{center}
            \begin{tabular}{||c||c||c||c||c||c||c||c||c||}
            \hline
             $Re_\tau$ & $180$ & $395$ & $590$ & $1000$ & $5200$ & $10^4$ & $10^5$ & $10^6$ \\
            \hline
                $C_\Pi$  & 5.93 & 2.68 & 2.25 & 1.92 & 1.69 & 1.68 & 1.68 & 1.68 \\
                \rule{0pt}{2ex}
                $C_s$  & 0.78 & 0.78 & 0.78 & 0.78 & 0.78 & 0.78 & 0.78 & 0.78 \\
                \rule{0pt}{2ex}
                $C_{uv}$ & 0.407 & 0.540 & 0.563 & 0.600 & 0.625 & 0.625 & 0.625 & 0.625 \\
                \rule{0pt}{2ex}
                $\Pi_{\text{ref}}$ & 0.344 & 0.203 & 0.162 & 0.162 & 0.162 & 0.162 & 0.162 & 0.162 \\
                \rule{0pt}{2ex}
                $B$ & 4.68 & 4.97 & 5.07 & 5.07 & 5.07 & 5.07 & 5.07 & 5.07 \\
            \hline
        \end{tabular}
        \label{tab:tune_params}
    \end{center}
\end{table}

For Reynolds numbers larger than those with available DNS data, the parameters from $Re_\tau = 5200$ are used. The only exception is that $C_\Pi$ is adjusted to close the AMI equation as the viscous term continues to decrease toward zero with increasing $Re_\tau$ (a very small effect). To verify the successful selection of model parameters according to the above procedure, table \ref{tab:stat_verification} shows the target DNS values and the TRIS results. The average zeroth and first moment of streamwise velocity increases with increasing $Re_\tau$ because the flow variables are normalized by friction velocity and length scales. The statistics in table \ref{tab:stat_verification} and the remainder of Section \ref{sec:results} are calculated over $128$ large-eddy turnover times, $h / u_\tau$.

Importantly, it is demonstrated here that the theoretical resolution of $35-40\%$ of the Reynolds shear stress can be achieved with the present TRIS formulation, however simple it may be.
Without significant effort to optimize the computational runtime, a flow through time for a domain of length $L_x = 8\pi$ takes $\sim 1$ minute on a single processor with a  desktop computer.
Furthermore, simulations up to $Re_\tau=10^6$ were performed without increase in computational cost. Note that these results are for a specific domain size and grid resolution, and that the $C_\Pi$ coefficient require retuning for different choices of grid spacing and domain size.

\begin{table}
    \captionsetup{justification=raggedright, singlelinecheck=false,width=\textwidth}
    \caption{Verification of TRIS implementation and parameter selection for reproducing target values from DNS.}
    \vspace{2mm}
    \begin{center}
            \begin{tabular}{||c||c | c||c | c||c | c||c | c || c | c||c||c||c||}
             $Re_\tau$ & \multicolumn{2}{c||}{$180$} & \multicolumn{2}{c||}{$395$} & \multicolumn{2}{c||}{$590$} & \multicolumn{2}{c||}{$1000$} & \multicolumn{2}{c||}{$5200$} & $10^4$ & $10^5$ & $10^6$ \\
            ~ & DNS & TRIS & DNS & TRIS & DNS & TRIS & DNS & TRIS & DNS & TRIS & TRIS & TRIS & TRIS \\
            \hline
            \hline
                \rule{0pt}{2.5ex}
                $\overline{\left<\widetilde{u}\right>_0}$ & 15.7 & 15.6 & 17.6 & 17.5 & 18.6 & 18.5 & 20.0 & 19.8 & 24.1 & 23.8 & 25.4 & 31.0 & 36.6 \\
                \rule{0pt}{2ex}
                $\overline{\left<\widetilde{u}\right>_1}$ & 17.3 & 17.2 & 19.0 & 18.9 & 20.0 & 19.9 & 21.3 & 21.2 & 25.4 & 25.2 & 26.8 & 32.4 & 38.0 \\
            \hline
                \rule{0pt}{2ex}
                $\frac{C_{f,\text{res}}}{C_f}$ & 0.38 & 0.38 & 0.36 & 0.36 & 0.37 & 0.37 & 0.34 & 0.35 & 0.39 & 0.36 & 0.36 & 0.39 & 0.37 \\
                \rule{0pt}{2ex}
                $\frac{C_{f,\text{unres}}}{C_f}$ & 0.41 & 0.41 & 0.54 & 0.54 & 0.56 & 0.56 & 0.61 & 0.60 & 0.59 & 0.63 & 0.63 & 0.63 & 0.63 \\
                \rule{0pt}{2ex}
                $\frac{C_{f,\text{lam}}}{C_f}$ & 0.21 & 0.21 & 0.10 & 0.10 & 0.07 & 0.07 & 0.05 & 0.05 & 0.01 & 0.01 & 0.01 & 0.00 & 0.00 \\
            \hline
        \end{tabular}
        \label{tab:stat_verification}
    \end{center}
\end{table}

\begin{figure}
    \centerline{\includegraphics[width=0.95\linewidth]{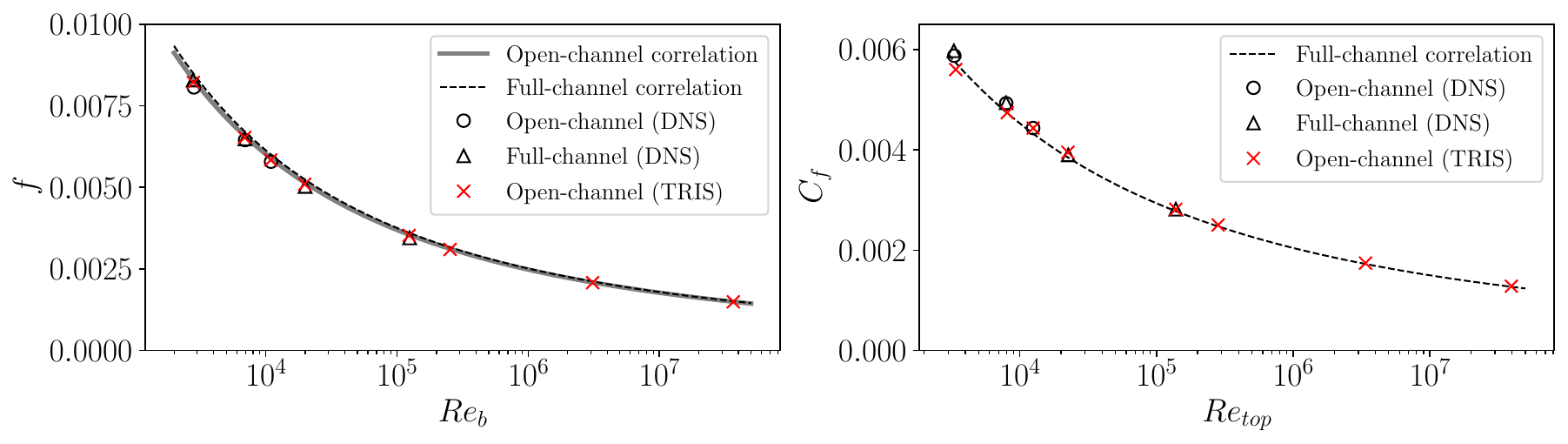}}
    \captionsetup{justification=raggedright, singlelinecheck=false,width=\textwidth}
    \caption[friction]{Friction factor plotted against the bulk Reynolds number (left) and skin friction plotted against the Reynolds number based on $\overline{u}_\text{top}$ (right). The dashed lines plot full-channel correlations \citep{Dean1978}. The grey solid line plots the friction factor correlation for an open-channel flow \citep{Bellos2018}.}
    \label{fig:friction_factor}
\end{figure}

The validity of model parameter extrapolation to high Reynolds number is verified by inspecting TRIS results for the friction factor, $f=2/\langle \overline{u} \rangle_0^2$, as a function of bulk Reynolds number, $Re_b=Re_\tau\langle\overline{u}\rangle_0$, in figure\ \ref{fig:friction_factor}. Here, the friction factor is compared with the correlation from \cite{Dean1978} (dashed lines). The successful alignment of TRIS with DNS up to $Re_\tau = 5200$ and with the empirical correlation at all $Re_\tau$ demonstrates the success and robustness of the choice of parameters in table \ref{tab:tune_params}.

\subsection{Flow Structure}

\begin{figure}
    \centerline{\includegraphics[width=1.0\linewidth]{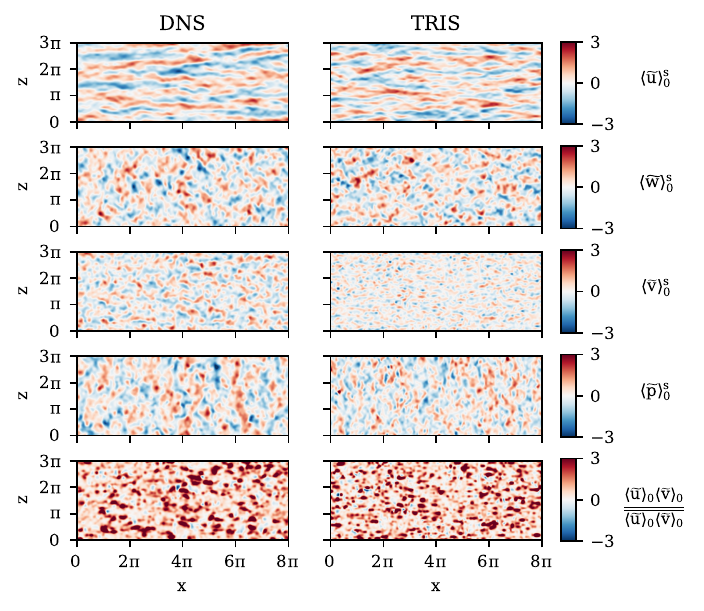}}
    \captionsetup{justification=raggedright, singlelinecheck=false,width=\textwidth}
    \caption{Instantaneous snapshots of the standardized (denoted by a superscript $s$) $\left<\widetilde{u}\right>_0$, $\left<\widetilde{w}\right>_0$, $\left<\widetilde{v}\right>_0$, and $\left<\widetilde{p}\right>_0$ fields in decending order at $Re_\tau = 395$. The covariance field, $\left<\widetilde{u}\right>_0\left<\widetilde{v}\right>_0$, is normalized by its mean}. Snapshots are based on the field imposed by a spectral cutoff filter of $k_\text{cut} h=16$ for DNS (left column) to match the grid resolution of TRIS (right column). Videos of the temporal evolution of these fields are available in Supplementary Material. For TRIS specifically, a Python code running the time progression of these fields through Jupyter notebook is available at \url{https://cocalc.com/share/public_paths/fa88e6bf9eea2307452e4f69f0a3bf7f8c65d8bd/figure-4}.
    \label{fig:395_snapshot}
\end{figure}

The flow structure in TRIS is now inspected by comparison to the DNS data. For an apples-to-apples comparison, only the open-channel flow DNS data ($180 \leq Re_\tau \leq 590$) are considered in this subsection. First, an instantaneous snapshot from open-channel DNS is integrated in the wall-normal direction and filtered using the spectral cutoff filter corresponding to the TRIS grid resolution.
The TRIS simulation reaches a statistically stationary state with streamwise-oriented streaky structures that exhibit self-sustaining dynamics, as shown in the top row of figure\ \ref{fig:395_snapshot}. The TRIS results on the right-side column are in comparison with (filtered) zeroth moment fields from DNS on the left-side column. 
Results at $Re_\tau = 590$ are visually similar to the $Re_\tau = 395$ snapshots shown here.

In order to emphasize flow structure, all fields in figure\ \ref{fig:395_snapshot} are standardized (denoted with a superscript $s$), i.e., fluctuations normalized by their standard deviation.
While $\langle \widetilde{u} \rangle_0^s$ (top row) exhibit relatively realistic streaky structure, $\langle \widetilde{w} \rangle_0^s$ and $\langle \widetilde{v} \rangle_0^s$ (second and third row, respectively) do not show similar streaks in the TRIS results, in agreement with the flow structure observed from the DNS results.
Similarly, $\langle \widetilde{p} \rangle_0^s$ and $\left<\widetilde{u}\right>_0\left<\widetilde{v}\right>_0$ (fourth and last row, respectively) do not contain streamwise-oriented streaks. 
These TRIS results demonstrate that the 2D-based integral moment equations, derived from the Navier-Stokes equations, are sufficient to generate self-sustaining turbulence with qualitatively realistic structure in a 2D/3C representation.

\begin{figure}
    \centering
    \includegraphics[trim={0 35 0 0}, clip, width=5in]{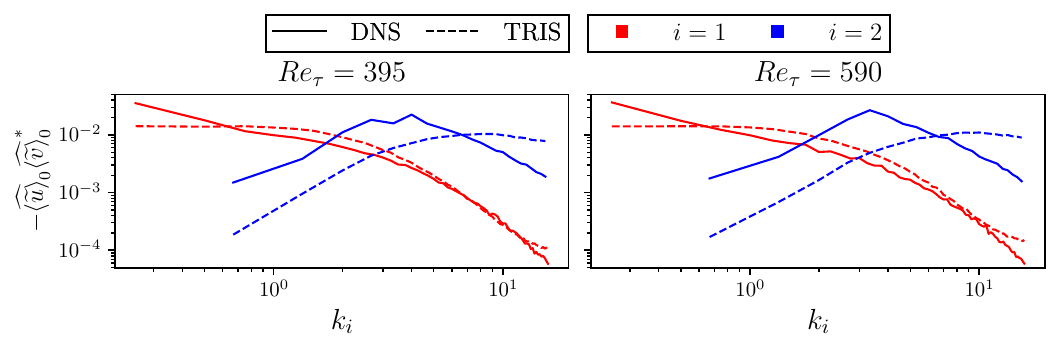} \\
    \includegraphics[trim={0 35 0 42}, clip,width=5in]{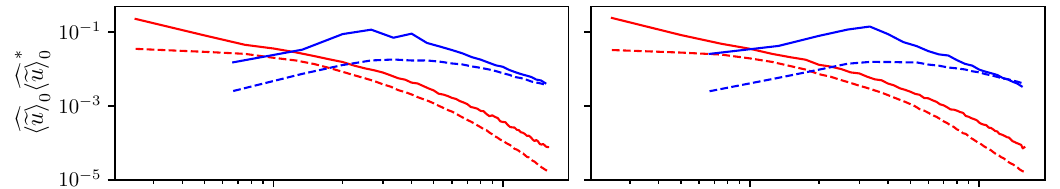} \\
    \includegraphics[trim={0 35 0 42}, clip,width=5in]{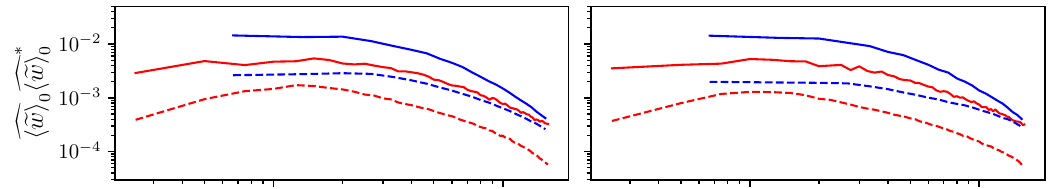} \\
    \includegraphics[trim={0 35 0 42}, clip,width=5in]{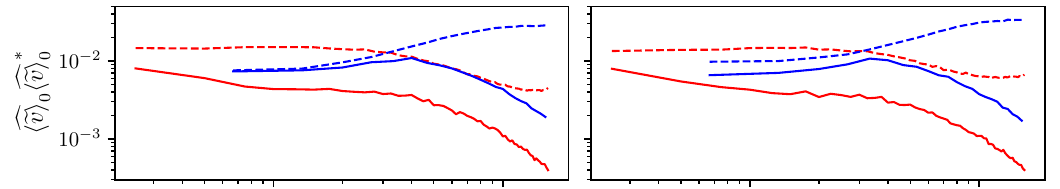} \\
    \includegraphics[trim={0 0 0 42}, clip,width=5in]{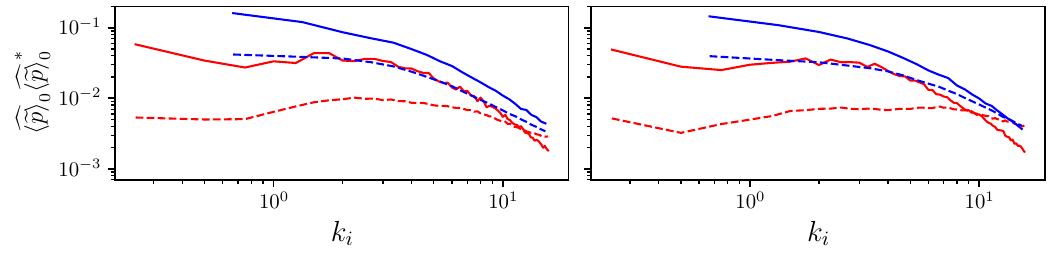}
    \captionsetup{justification=raggedright, singlelinecheck=false,width=\textwidth}
    \caption{Streamwise (red) and spanwise (blue) spectral distributions of the resolved shear, streamwise, spanwise, and wall-normal Reynolds stress components and resolved pressure in descending order at $Re_\tau=395$ (left) and $Re_\tau=590$ (right). The Fourier transform, $\widehat{\phi}$ of the resolved component is multiplied by its complex conjugate, $\widehat{\phi}^*$. The solid and dashed lines represent DNS and TRIS, respectively, and $k_i$ is non-dimensionalized by the height of the open-channel, $h$.}
    \label{fig:spectral_comparison}
\end{figure}

For a quantitative evaluation of the present TRIS formulation (in terms of results that are not set or tuned via closure parameter manipulation), the streamwise and spanwise (co-)spectra of the Reynolds shear stress and three kinetic energy components are shown in figure\ \ref{fig:spectral_comparison} for $Re_\tau = 395$ and $Re_\tau = 590$. Here, $k_i$ is non-dimensionalized by the height of the open-channel, $h$.
The top row shows the co-spectra for the Reynolds shear stress. The sum of the co-spectra over all streamwise ($k_1 = k_x$) or spanwise ($k_2 = k_z$) modes is the zeroth-moment of the Reynolds shear stress as it shows up in the AMI balance, \eqref{eq:AMI}, which matches the DNS by means of parameter tuning. More interestingly, the TRIS results show a good degree of success in replicating the shape of the distribution of the Reynolds shear stress as a function of both streamwise (red) and spanwise (blue) wavenumber. In keeping with the structure observed in figure\ \ref{fig:395_snapshot}, the co-spectrum peaks at the lowest streamwise wavenumber and at an intermediate spanwise wavenumber. The TRIS results thus reproduce this basic structure, although the spectrum peaks at a larger spanwise wavenumber compared to DNS, which is related to the observation from figure\ \ref{fig:395_snapshot} that the typical streak width is generally under-predicted by the current TRIS formulation.
The TRIS and DNS results do not show significant sensitivity to $Re_\tau$.

The second to fourth rows of figure\ \ref{fig:spectral_comparison} show the spectra for each of the three components of kinetic energy: streamwise, spanwise, and wall-normal, respectively. The shape of the spectra of the streamwise and spanwise velocity components produced by the present TRIS formulation are generally similar to the DNS results, but with lower overall magnitude. That is, the root-mean-square of $\langle \widetilde{u} \rangle_0$ and $\langle \widetilde{w} \rangle_0$ 
are under-predicted by TRIS. 
For the wall-normal velocity component, the shape of the TRIS spectra with respect to $k_1$ is relatively accurate. However, TRIS shows a peak at the highest resolved $k_2$, however, which is significantly different than the DNS spectrum.
As a result, the $\langle \widetilde{v} \rangle_0$ root-mean-square shows an over-prediction by TRIS.
The last row of figure\ \ref{fig:spectral_comparison} illustrates the resolved pressure spectra. Here, the TRIS pressure spectra are relatively realistic, though not perfect, with an indication that the magnitude of the large-scale pressure fluctuations are under-predicted.

\begin{figure}
  \centering
  \begin{subfigure}[b]{0.49\linewidth}
    \includegraphics[width=\linewidth]{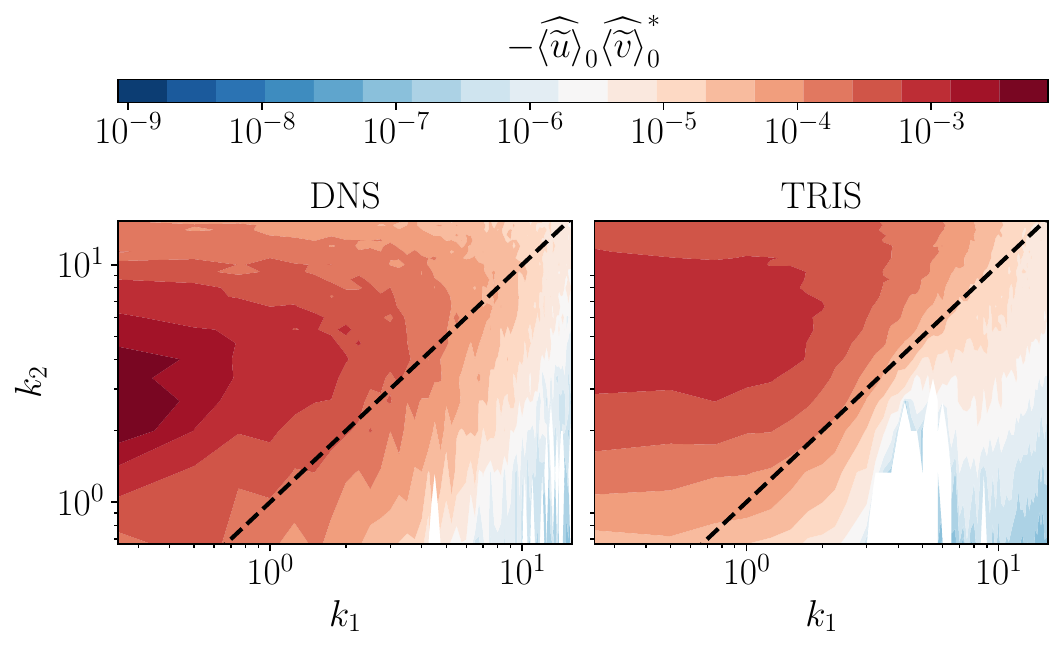}
    \caption{}
    \label{fig:sub1}
  \end{subfigure}
  \hfill
  \begin{subfigure}[b]{0.49\linewidth}
    \includegraphics[width=\linewidth]{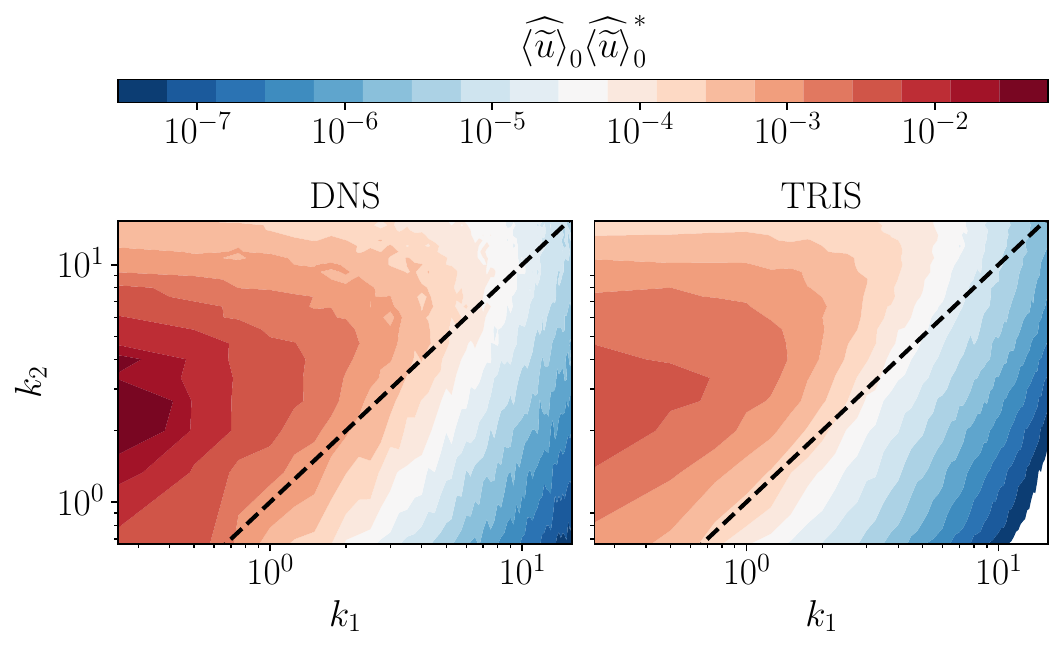}
    \caption{}
    \label{fig:sub2}
  \end{subfigure}

  \begin{subfigure}[b]{0.49\linewidth}
    \includegraphics[width=\linewidth]{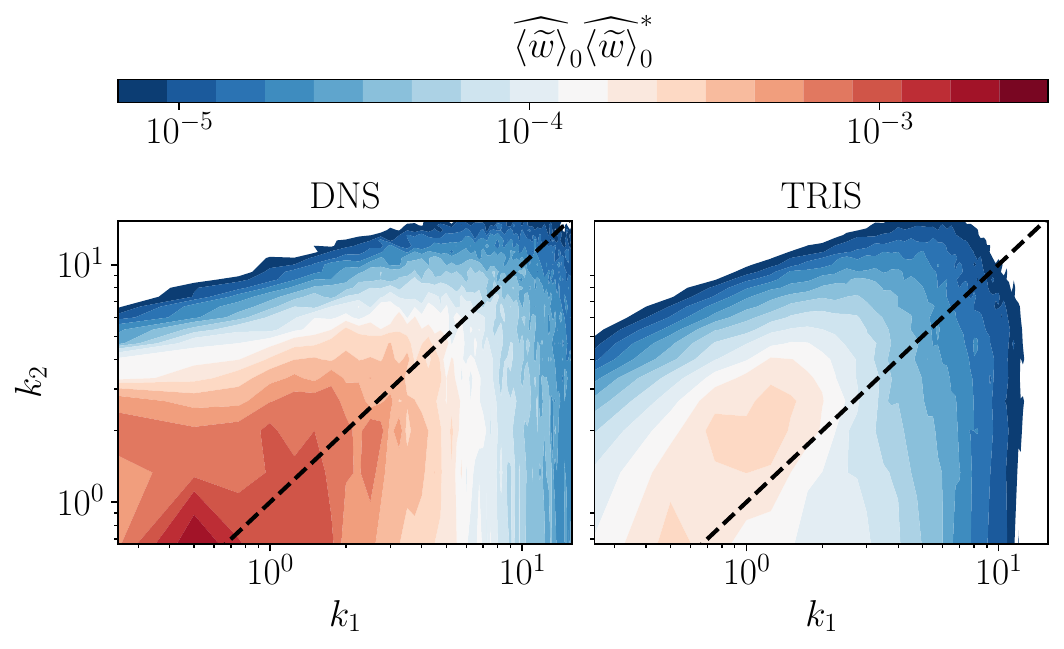}
    \caption{}
    \label{fig:sub1}
  \end{subfigure}
  \hfill
  \begin{subfigure}[b]{0.49\linewidth}
    \includegraphics[width=\linewidth]{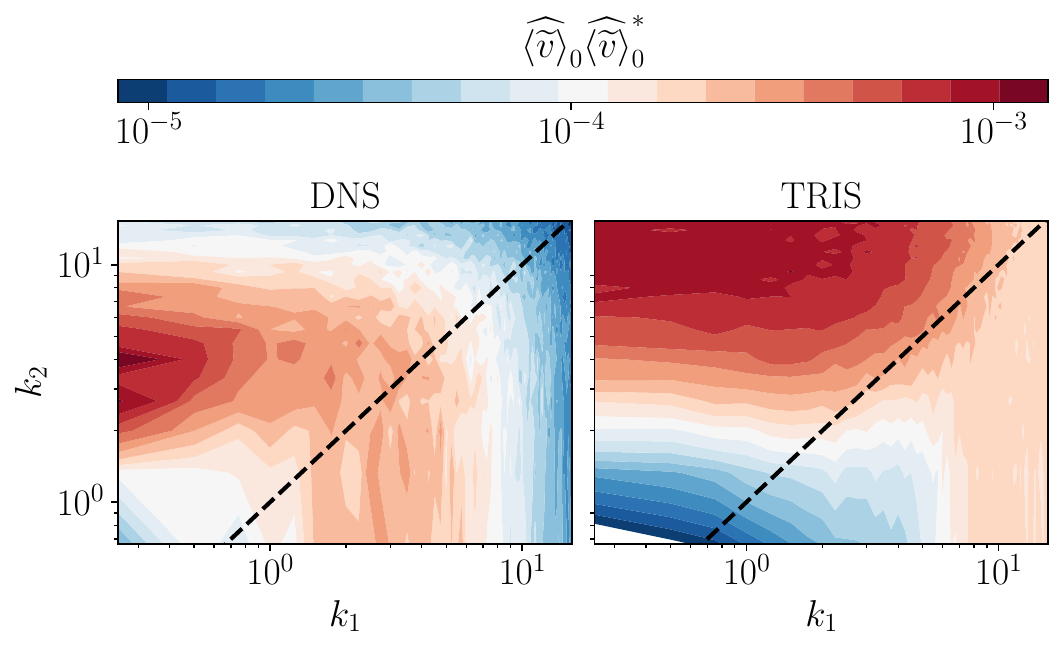}
    \caption{}
    \label{fig:sub2}
  \end{subfigure}

  \begin{subfigure}[b]{0.49\linewidth}
    \includegraphics[width=\linewidth]{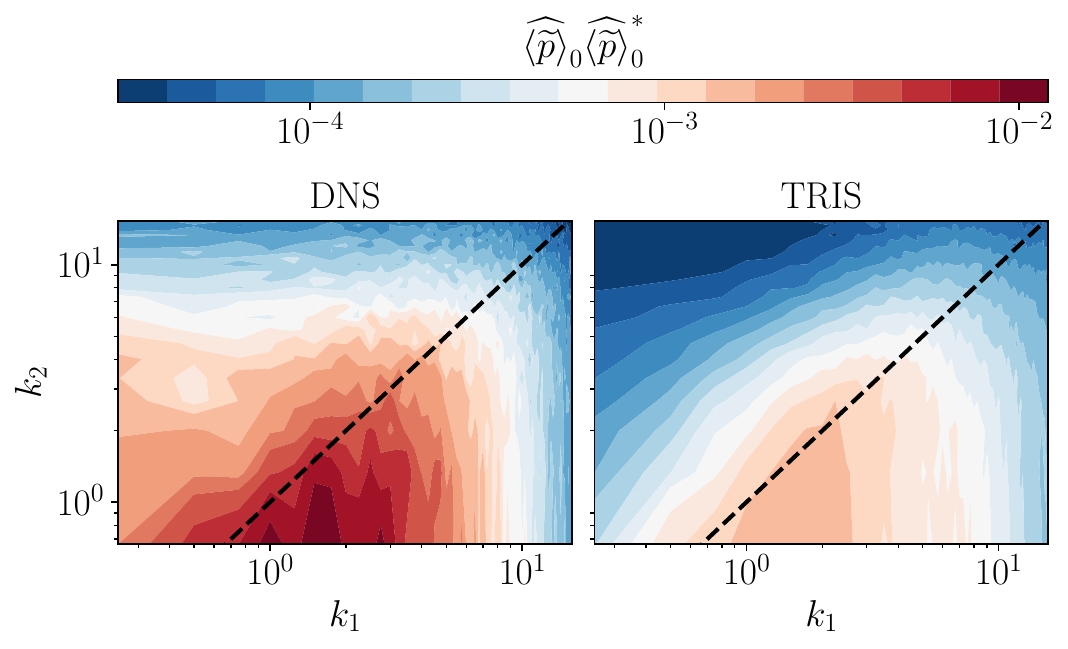}
    \caption{}
    \label{fig:sub2}
  \end{subfigure}

  \captionsetup{justification=raggedright, singlelinecheck=false,width=\textwidth}
  \caption{2D spectral distribution of the resolved Reynolds shear stress, streamwise, spanwise, and wall-normal variances, and pressure across subplots (a)-(e), respectively. Results are plotted at $Re_\tau=395$ and the Fourier transform, $\widehat{\phi}$, of the resolved component is multiplied by its complex conjugate, $\widehat{\phi}^*$. The dashed black line is a linear line with a slope of unity and a vertical intercept of zero ($k_2=k_1$). In each subplot, the spectral fields of DNS and TRIS are on the left and right, respectively. Streamwise ($k_1$) and spanwise ($k_2$) are non-dimensionalized by the height of the open-channel, $h$.}
  \label{fig:2D_spectral_comparison}
\end{figure}

Two-dimensional spectral comparisons between TRIS and DNS at $Re_\tau = 395$ are also illustrated in figure\ \ref{fig:2D_spectral_comparison}, providing a holistic view on the resolved components of the velocity variances/covariances and resolved pressure. The black dashed line, $k_1 = k_2$, separates predominantly streamwise-oriented modes (upper left corner) from predominantly spanwise-oriented modes (bottom right corner). TRIS correctly reproduces the relative shape of the u-v co-spectrum with its bias toward streamwise-oriented structures, figure\ \ref{fig:2D_spectral_comparison}a. It is noted, however, that the co-spectrum maximum is slightly shifted to higher spanwise wavenumber in the TRIS results, in agreement with the 1D spectra shown above. 
TRIS also generally reproduces the shape of the wall-parallel (streamwise and spanwise) velocity spectra well, figure\ \ref{fig:2D_spectral_comparison}b,c. The streamwise velocity is dominated by streamwise-oriented modes, while the spanwise velocity is more isotropic. As already observed, the magnitudes of these TRIS spectra are under-predicted. 
As for the resolved wall-normal variance, the tendency of TRIS to over-predict the most active wavenumbers is again observed. However, the shape of the wall-normal velocity spectrum from TRIS is not altogether unrealistic (though significantly shifted to higher wavenumbers). 
Lastly, the resolved pressure variance shows that pressure fluctuations occur mostly at low spanwise and intermediate streamwise wavenumbers, corresponding to short and wide structures as illustrated in the fourth row of figure\ \ref{fig:395_snapshot}. TRIS also presents a similar behavior, but the overall magnitude is under-predicted, as previously observed.

\subsection{Single-point Statistics}


While the model parameters in table \ref{tab:tune_params} were selected specifically to cause TRIS to provide accurate statistics for the quantities shown in table \ref{tab:stat_verification}, table \ref{tab:stat_comparison} shows additional single-point statistics for TRIS and open-channel DNS to further probe the quantitative accuracy of the current TRIS implementation.
First, the root-mean-square values for the zeroth moment of the three velocity components and pressure are given. As evidenced in the above spectra, the wall-parallel velocity fluctuation magnitudes are under-predicted while the wall-normal magnitude is over-predicted. This trend is consistent across all wavenumbers for all open-channel flow Reynolds numbers, $180 \leq Re_\tau \leq 590$.
The under-prediction of the zeroth moment of pressure fluctuation magnitude may be related to the over-prediction of the wall-normal velocity fluctuations. The root-mean-square values for the first moments (not shown) have similar discrepancies between TRIS and DNS.

\begin{table}
    \captionsetup{justification=raggedright, singlelinecheck=false,width=\textwidth}
    \caption{Additional single-point statistics of TRIS and DNS at various $Re_\tau$: root-mean-square ($\text{RMS}\{\phi\}$), correlation coefficient ($r(\phi, \psi)$),
    skewness ($S\{\phi\}$), and excess kurtosis ($K\{\phi\}$) are listed in descending order. The last section of rows list the AMI balance (values also illustrated in figure\ \ref{fig:AMI_balance}). DNS data is available up to $Re_\tau=590$ while TRIS data is up to $Re_\tau=10^6$.}
    \vspace{2mm}
    \begin{center}
            \begin{tabular}{||c||c | c||c | c||c | c||c || c||c||c||}
             $Re_\tau$ & \multicolumn{2}{c||}{$180$} & \multicolumn{2}{c||}{$395$} & \multicolumn{2}{c||}{$590$} & $1000$ & $10^4$ & $10^5$ & $10^6$ \\
            ~ & DNS & TRIS & DNS & TRIS & DNS & TRIS & TRIS & TRIS & TRIS & TRIS \\
            \hline
                \rule{0pt}{2.5ex}
                $\text{RMS}\{\langle \widetilde{u} \rangle_0\}$ & 0.87 & 0.50 & 0.84 & 0.48 & 0.88 & 0.46 & 0.46 & 0.47 & 0.48 & 0.48 \\
                \rule{0pt}{2ex}
                $\text{RMS}\{\langle \widetilde{v} \rangle_0\}$ & 0.38 & 0.72 & 0.37 & 0.71 & 0.36 & 0.76 & 0.74 & 0.79 & 0.80 & 0.80 \\
                \rule{0pt}{2ex}
                $\text{RMS}\{\langle \widetilde{w} \rangle_0\}$ & 0.35 & 0.22 & 0.34 & 0.17 & 0.33 & 0.15 & 0.15 & 0.15 & 0.15 & 0.15 \\
                \rule{0pt}{2ex}
                $\text{RMS}\{\langle \widetilde{p} \rangle_0\}$ & 1.01 & 0.79 & 0.97 & 0.61 & 0.93 & 0.62 & 0.61 & 0.66 & 0.67 & 0.68 \\
                \rule{0pt}{2ex}
                $\text{RMS}\{\langle \widetilde{u} \rangle_0\langle \widetilde{v} \rangle_0\}$ & 0.38 & 0.43 & 0.35 & 0.38 & 0.36 & 0.39 & 0.38 & 0.40 & 0.41 & 0.42 \\
                
            \hline
                \rule{0pt}{2.5ex}
                $r(\left<\widetilde{u}\right>_0,\left<\widetilde{v}\right>_0)$ & -0.58 & -0.53 & -0.57 & -0.53 & -0.58 & -0.53 & -0.52 & -0.49 & -0.48 & -0.48 \\

            \hline
                \rule{0pt}{2.5ex}
                $S \{\left<\widetilde{u}\right>_0\}$ & -0.14 & 0.21 & -0.16 & 0.039 & -0.19 & 0.011 & 0.019 & 0.032 & 0.038 & 0.050 \\
                \rule{0pt}{2ex}
                $S \{\left<\widetilde{v}\right>_0\}$ & 0.22 & -0.18 & 0.16 & -0.59 & 0.18 & -0.79 & -0.73 & -0.66 & -0.65 & -0.64 \\
                \rule{0pt}{2ex}
                $S \{\left<\widetilde{p}\right>_0\}$ & -0.21 & 0.25 & -0.11 & 0.14 & -0.076 & 0.096 & 0.082 & 0.072 & 0.060 & 0.072 \\
                \rule{0pt}{2ex}
                $S \{\langle \widetilde{u} \rangle_0\langle \widetilde{v} \rangle_0\}$ & -2.40 & -2.92 & -2.28 & -2.74 & -2.31 & -2.72 & -2.65 & -2.56 & -2.48 & -2.55 \\
            \hline
                \rule{0pt}{2.5ex}
                $K \{\left<\widetilde{u}\right>_0\}$ & -0.19 & -0.050 & -0.19 & -0.23 & -0.23 & -0.31 & -0.28 & -0.26 & -0.27 & -0.25 \\
                \rule{0pt}{2ex}
                $K \{\left<\widetilde{v}\right>_0\}$ & 0.095 & 1.4 & 0.022 & 1.7 & 0.040 & 1.7 & 1.6 & 1.28 & 1.23 & 1.2 \\
                \rule{0pt}{2ex}
                $K \{\left<\widetilde{p}\right>_0\}$ & 0.50 & 1.23 & 0.47 & 0.97 & 0.42 & 1.25 & 1.11 & 1.01 & 1.00 & 1.00 \\
                \rule{0pt}{2ex}
                $K \{\langle \widetilde{u} \rangle_0\langle \widetilde{v} \rangle_0\}$ & 9.82 & 16.42 & 8.67 & 14.92 & 8.97 & 14.18 & 13.50 & 13.07 & 12.24 & 13.41 \\
        \end{tabular}
        \label{tab:stat_comparison}
    \end{center}
\end{table}

\begin{figure}
    \centerline{\includegraphics[width=0.95\linewidth]{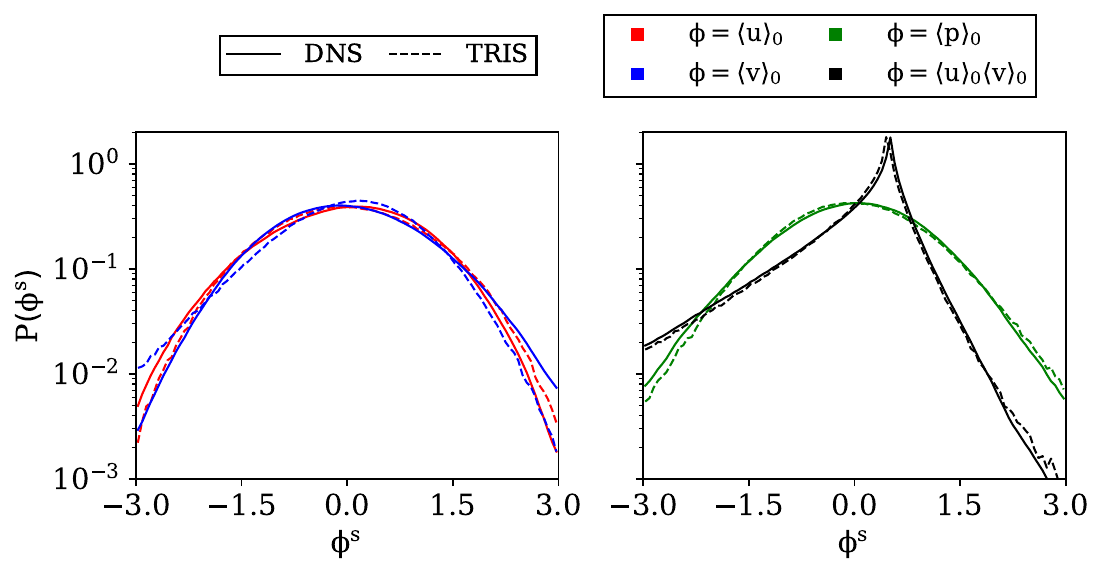}}
    \captionsetup{justification=raggedright, singlelinecheck=false,width=\textwidth}
    \caption{Standardized (denoted with superscript ``s'') probability density functions of the zeroth moments of the streamwise and wall-normal velocity (left) and zeroth moment of pressure and resolved shear stress (right). Solid lines correspond to DNS and the dashed lines correspond to TRIS and comparisons are made for $Re_\tau=395$.}
    \label{fig:pdf}
\end{figure}

DNS shows that the zeroth moment of the streamwise and wall-normal velocities have slight negative and positive skewness, respectively. (The skewness of the spanwise velocity is zero due to symmetry.) Meanwhile, their excess kurtosis is also close to zero, indicating small departures from Gaussianity. The probability density function (PDF) of each of these velocity components is shown in figure\ \ref{fig:pdf} (left). The TRIS results, meanwhile, show a very small positive skewness for streamwise velocity and a larger negative skewness and positive excess kurtosis for the wall-normal velocity. Nonetheless, the TRIS PDFs appear relatively close to the Gaussian shape, so the discrepancy with DNS is not strong.
The zeroth moment of pressure in DNS has a slight negative skewness and mild positive excess kurtosis. TRIS predicts a slight positive skewness with a larger excess kurtosis. The pressure PDFs are compared in figure\ \ref{fig:pdf} (right).

The root-mean-square of the product $\langle \widetilde{u} \rangle_0\langle \widetilde{v} \rangle_0$ is relatively accurate in TRIS, while the negative correlation coefficient of $\langle \widetilde{u} \rangle_0$ and $\langle \widetilde{v} \rangle_0$ is relatively well-represented but under-predicted in magnitude. The PDF of $\langle \widetilde{u} \rangle_0\langle \widetilde{v} \rangle_0$ is shown in figure\ \ref{fig:pdf} (right). The skewness of $\langle \widetilde{u} \rangle_0\langle \widetilde{v} \rangle_0$ is strongly negative, which TRIS predicts quite well, though TRIS over-predicts its excess kurtosis.


The statistical results of these higher Reynolds number simulations ($1000 \varleq Re_\tau \varleq 10^6$) are shown in the last four columns of table \ref{tab:stat_comparison}. The root-mean-square of $\left< \widetilde{u} \right>_0$ and $\left< \widetilde{v} \right>_0$ display an increasing trend with respect to $Re_\tau$. Beyond $Re_\tau \sim 1000$, the TRIS predictions show little variation in the single-point statistics as Reynolds number increases.

\section{Concluding Discussion}\label{sec:conclusion}
This paper introduces a framework for Turbulence-Resolving Integral Simulations (TRIS) of wall-bounded flows. A proof-of-concept demonstration is shown for an open-channel configuration using instantaneous moment-of-momentum integral equations (derived from first principles) with closures based on an assumed profile.
The use of zeroth- and first-moment integral equations in a 2D (streamwise-spanwise) domain provides a sufficient basis for reproducing the self-sustaining process of large-scale streaks in wall-bounded turbulence.
The resulting 2D/3C TRIS simulations yield a qualitatively realistic structure for the three velocity components and pressure fields.
With wall-parallel resolution of $h/5$, DNS evidence suggests that the approach can directly resolve $35 - 40\%$ of the Reynolds shear stress responsible for turbulent skin friction enhancement. This estimate is relatively constant across a wide range of Reynolds numbers in open-channel and full-channel DNS, $180 \leq Re_\tau \leq 5200$, leading the authors to speculate that such resolution will hold for much larger Reynolds numbers.
The cost of TRIS is very low compared to established turbulence-resolving techniques such as LES and DNS, with one flow-through time taking $\sim 1$ minute on a single processor for a channel flow, even at very large Reynolds number,
using an unoptimized Python code.


The TRIS framework allows for apples-to-apples quantitative comparisons with DNS data (or experimental data, if available), allowing for a detailed analysis of model accuracy.
Overall, the comparisons of spectra and single-point statistics underscore some areas of accuracy for the present TRIS closure model while also highlighting some deficiencies. In particular, the present closures allow for an unrealistically large magnitude of high wavenumber wall-normal velocity fluctuations.
The authors speculate that more accurate closure models, e.g., for $\widetilde{p}_{\text{top}} - \widetilde{p}_{\text{bot}}$, could help TRIS produce a more accurate spectra of wall-normal fluctuations with respect to spanwise wavenumber. This could in turn also help yield a more accurate distribution of kinetic energy between the three components. Further work developing physics-based models is deferred to future work.

A related quasi-2D/3C approach to reduced-order modeling of self-sustaining wall-bounded turbulence is the restricted nonlinear (RNL) model \citep{Thomas2014}, which resolves the flow in the spanwise and wall-normal directions while severely restricting the representation of streamwise variations. In comparison, the TRIS approach is well suited for extension to a more general class of flows that are not periodic in the streamwise (or spanwise) direction.

A number of interesting extensions are possible. A multi-layer approach to TRIS could be developed based on performing wall-normal integrals with respect to the inner, log, and outer layer, which can potentially capture more physics at the expense of higher computational cost. 
Of course, identification of the region boundaries and specifying appropriate interface conditions will require detailed study. Analysis by \cite{Kwon2021} on DNS of an isolated logarithmic region could provide instrumental insight on modeling ideas for this approach.
One potentially fruitful topic for future investigation is a more detailed analysis of the production and transport of turbulent kinetic energy from the perspective of instantaneous wall-normal integrals. This could provide more insight into the shortcomings of the present closures and potential pathways for developing closures with higher physical fidelity.

The ability of TRIS to resolve large-scale motions, which have important sensitivities to favorable and adverse pressure gradients, motivates future development targeting engineering-relevant flows.
Future work will aim to formulate the TRIS equations for (external) boundary layer flows. The wall-normal integral of velocity diverges for a semi-infinite domain, so the formulation for boundary layers should be done in terms of velocity defect relative to an irrotational outer flow solution. The AMI equation for boundary layers is formulated this way \citep{Elnahhas_2022}. The equations governing the instantaneous zeroth and first moment of the velocity defect can be formulated and the freestream pressure gradient term would be formally closed. Lacking a no-penetration upper boundary, the boundary layer TRIS formulation will need to account for interaction with an irrotational freestream flow with zero/nonzero-pressure gradients and boundary layer induced fluctuations. 
Also, the streamwise growth of a boundary layer (i.e., lack of streamwise periodicity) will necessitate the development of realistic inflow boundary conditions, which could be based on recycling-rescaling concepts used in LES and DNS \citep{Lund1998,Spalart2006}.
We have avoided such complications in the present formulation in order to focus on the proof of concept for TRIS itself in terms of self-sustaining dynamics. The $35-40\%$ resolution of the Reynolds shear stress integral by TRIS shown in figure\ \ref{fig:AMI_balance} will also need to be reassessed using DNS of spatially developed boundary layers, though the authors expect any changes to be minor.

Importantly, a truly predictive approach (for engineering quantities of interest) requires more work to establish physics-based closure models to improve the accuracy and general applicability of TRIS compared to the present proof of concept.



\section*{Acknowledgments}

TR was supported by the Department of Defense (DoD) National Defense Science \& Engineering Graduate (NDSEG) Fellowship Program. MW and SS were supported by the Air Force Office of Scientific Research under award number FA9550-24-1-0127. PJ was supported by the National Science Foundation under CAREER award No. 2340121.

\section*{Declaration of Interests.}

The authors report no conflict of interest.

\bibliographystyle{jfm}
\bibliography{jfm}

\newpage

\appendix

\section{Verification and Statistical Convergence of Direct Numerical Simulations}\label{sec:DNSverification}

For the lower friction Reynolds number flows ($Re_\tau = 180, 395, 590$), the Navier-Stokes equations are solved on a staggered Cartesian grid using a second-order central difference scheme in the wall-parallel directions and an explicit third-order Runge-Kutta scheme for time advancement \citep{Lozano2018}. Full-channel flow simulations were executed and compared against \citet{Moser1999} to ensure proper spatial discretization for the open-channel flow application. These lower Reynolds number simulations were ran out for twenty large-eddy turnover times. For the higher friction Reynolds number flows ($Re_\tau = 1000, 5200$), velocity data was gathered from John Hopkins Turbulence Database \citep{Graham2016, Lee2015}, where the solver uses a Fourier-Galerkin pseudo-spectral method for the wall-parallel directions and a third-order Runge-Kutta scheme for time advancement. These simulations were ran out for roughly one large-eddy turnover time.

\begin{figure}
    \centering
    \includegraphics[width=0.85\linewidth]{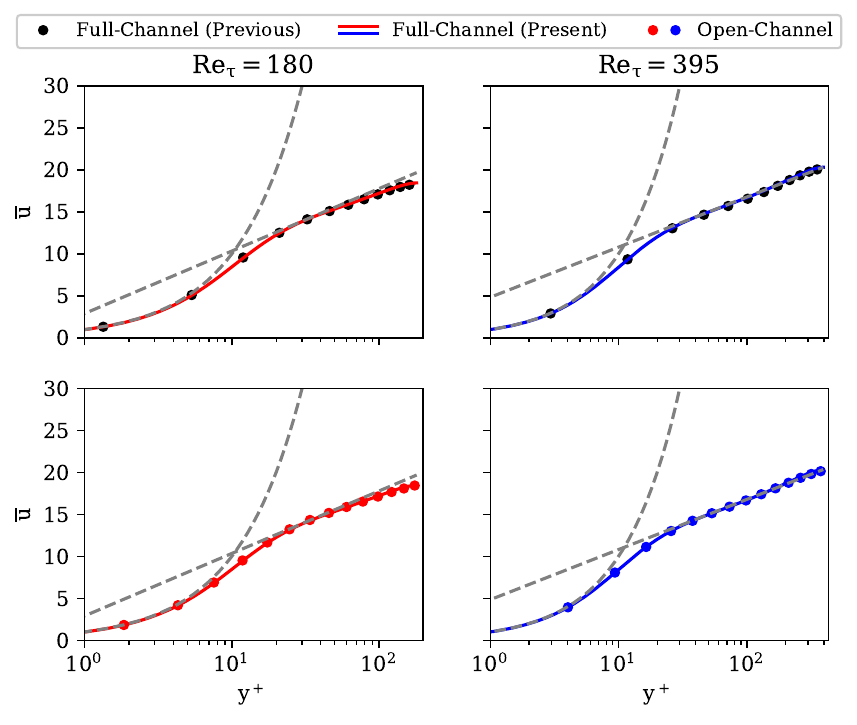}
    \captionsetup{justification=raggedright, singlelinecheck=false,width=\textwidth}
    \caption{Reynolds averaged mean velocity profiles of channel flows at $Re_\tau=180$ (left column) and $Re_\tau=395$ (right column). The ``Full-Channel (Previous)'' label (black circular markers) corresponds to the profiles gathered from \cite{Moser1999} whereas the ``Full-Channel (Present)'' and ``Open-Channel'' labels (colored lines and circular markers, respectively) correspond to the author's DNS data.}
    \label{fig:mean_profiles}
\end{figure}

\begin{figure}
    \centering
    \includegraphics[width=0.85\linewidth]{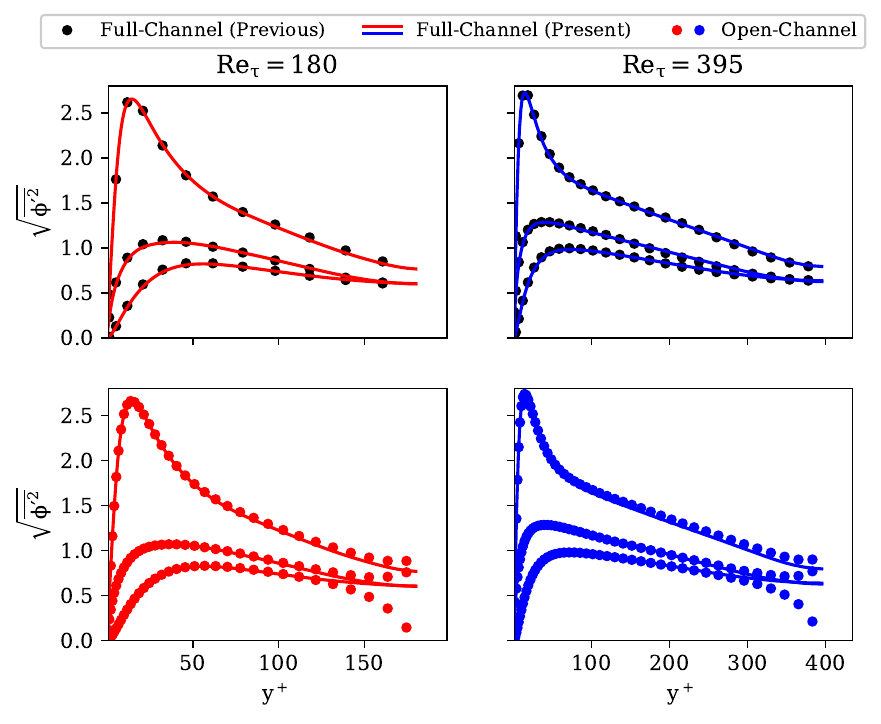}
    \captionsetup{justification=raggedright, singlelinecheck=false,width=\textwidth}
    \caption{Root-mean-square profiles of channel flows at $Re_\tau=180,395$ (left and right columns, respectively). ``Full-Channel (Previous)'', denoted by the black circular markers, corresponds to the profiles gathered from \cite{Moser1999} while ``Full-Channel (Present)'' and ``Open-Channel'' (denoted by colored lines and circular markers, respectively) correspond to the author's DNS data. In all sub-plots, the streamwise, spanwise, and wall-normal components are distributed in descending order.}
    \label{fig:rms_profiles}
\end{figure}

To ensure the viability of DNS executed by the authors, statistical quantities of the mean velocity and root-mean-square profiles are compared against \cite{Moser1999}. The top row of figure\ \ref{fig:mean_profiles} illustrates that the current DNS simulations (at $Re_\tau=180, 395$) accurately captures mean velocity profiles for the full-channel flow. Here, the grey dashed lines correspond to the fits of the viscous sub-layer and log-layer region. 
The top row of figure\ \ref{fig:rms_profiles} further shows that, in the full-channel configuration, the root-mean-square statistics sufficiently matches with \cite{Moser1999}. This analysis demonstrates sufficient spatial discretization, providing confidence in the accuracy of DNS on the open-channel configuration for $180 \varleq Re_\tau \varleq 590$.

Further comparison of the full-channel and open-channel configurations are illustrated in the bottom rows of figure\ \ref{fig:mean_profiles} and \ref{fig:rms_profiles}. Here, the profile in the full-channel flow extends from the bottom wall (no-slip) to the centerline (no boundary condition) whereas the profile in the open-channel flow extends from the bottom wall (no-slip) to the top boundary (no-vorticity). Interestingly, the mean velocity profiles are statistically identical between these two configurations. However, notable discrepancies are observed in the root-mean-square velocity profiles for $Re_\tau y \gtrsim 125$, caused by the no-penetration condition at the top boundary. This boundary condition enforces the wall-normal fluctuations to be zero, causing the streamwise and spanwise fluctuations to increase at the top wall.

The statistical convergence of the resolved skin friction , $C_{f,\text{res}}$, in \eqref{eq:AMI} is further analyzed. Since large-scale motions (contributing to $C_{f,\text{res}}$) tend to have longer turnover times than smaller scales (contributing to $C_{f,\text{unres}}$), $C_{f,\text{unres}}$ is more statistically converged than its resolved counterpart. As used by \cite{Shirian2023}, the estimated statistical convergence error (or standard error of the mean) is computed with
\begin{equation}
    \phi_\text{error}=\frac{1}{N}\sqrt{\sum_{i=1}^N\left( \phi_{i,w} - \phi_m \right)^2},
    \label{eq:SEM}
\end{equation}
where $N$ is the number of time windows, $\phi_{i,w}$ is the average quantity across a particular time window, and $\phi_m$ is the statistically converged average quantity \citep{ross1998}. According to \cite{Shirian2023}, estimates of the statistical error can reasonably be computed across $N=4$ windows with a window length of $10$ turnover times on a full-channel flow with a domain size of $L_x=2\pi$ and $L_z=1\pi$. With the present DNS simulations running on a domain size of $L_x=8\pi$ and $L_z=3\pi$, which is 12 times larger than the previously mentioned domain size, the number of windows and window lengths are modified.

For the lower Reynolds number channel flow simulations ($Re_\tau=180,395,590$), \eqref{eq:SEM} of $C_{f,\text{res}}$ is computed across $N=20$ windows with window lengths of $1$ turnover time. The estimated statistical error of $C_{f,\text{res}}$ is less than $1.0\%$. Alternatively, for the higher Reynolds number simulations ($Re_\tau=1000, 5200$), the error of $C_{f,\text{res}}$ is computed across $N=2$ windows with window lengths of about $1$ turnover time. It turns out that the maximum $\phi_\text{error}$ value of $C_{f,\text{res}}$ is approximately $2.0\%$.

Higher confidence of the estimated error of the averaged $C_{f,\text{res}}$ is placed in the lower $Re_\tau$ values since more windows are available to compute over. The corresponding estimated values for higher $Re_\tau$ are (very) rough approximations since more velocity data is required to accurately compute the absolute average ($\phi_m$) of the resolved skin friction by turbulent enhancement (i.e., more time windows are required).


\section{Detailed Derivation of Closure Models for TRIS}\label{sec:derive_model}
For demonstrative purposes, a simple closure is presented by assuming that the skewed Coles wake velocity profile, \eqref{eq:coles_profile},
\begin{equation*}
    U_i = \left[ \frac{1}{\kappa} \text{ln} y + \left( \frac{1}{\kappa} \text{ln}Re_{\ast} + B \right) \right] e_{i,\ast} + \left[ \frac{2 \Pi}{\kappa} \text{sin}^2\left(\frac{\pi}{2}y\right)  \right] e_{i,\Pi},
\end{equation*}
captures the conditionally averaged wall-parallel profiles ($U_i(x_1,y, x_2,t) = \left\langle \widetilde{u_i} | \langle \widetilde{u_i} \rangle_0 , \langle \widetilde{u_i} \rangle_1 \right \rangle $). To capture the integral moments, \eqref{eq:moments-def} is applied on $U_i$, with the assumption that the moment integrals of the profile fluctuations about the conditional average ($\langle U_i^{\prime\prime} \rangle_{0}$ and $\langle U_i^{\prime\prime} \rangle_{1}$) are negligible,
\begin{equation}
    \langle \widetilde{u_i} \rangle_0 = \langle U_i \rangle_0 + \cancelto{\approx 0}{\langle U''_i \rangle_0},
	\hspace{0.2\linewidth} 
    \langle \widetilde{u_i} \rangle_1 = \langle U_i \rangle_1 + \cancelto{\approx 0}{\langle U''_i \rangle_1}.
    \label{eq:velocity decomposition}
\end{equation}
Applying \eqref{eq:moments-def} on \eqref{eq:coles_profile} generates the following relations,

\begin{equation}
    \langle U_i \rangle_0 
    = \left[ \frac{1}{\kappa}\left( \text{ln}Re_\ast-1 \right) + B \right] e_{i,\ast}
    + \frac{\Pi}{\kappa} e _{i,\Pi},
    \label{eq:Coles_mom0}
\end{equation}
\begin{equation}
    \langle U_i \rangle_1 
    = \left[ \frac{1}{\kappa}\left( \text{ln}Re_\ast-\frac{1}{2} \right) + B \right] e_{i,\ast}
    + \frac{\Pi}{\kappa} \left( 1+\frac{4}{\pi^2} \right) e _{i,\Pi},
\end{equation}
which are used to solve for $\text{ln}Re_\ast$, $e_{i,\ast}$, $\Pi$, and $e_{i,\Pi}$. Unit vectors, $e_{i,\ast}$ and $e_{i,\Pi}$ are defined such that 
\begin{equation}
    \begin{split}
        \tau_i = |\tau|
        \begin{cases}
            e_{1,\ast} = \text{cos}(\theta_\ast) & i=1 \\
            e_{2,\ast} = \text{sin}(\theta_\ast) & i=2
        \end{cases} \\
        \Pi_i = |\Pi|
        \begin{cases}
            e_{1,\Pi} = \text{cos}(\theta_\Pi) & i=1 \\
            e_{2,\Pi} = \text{sin}(\theta_\Pi) & i=2
        \end{cases}
    \end{split}
\end{equation}
where

\begin{equation}
    \begin{split}
    \theta_\ast = \text{tan}^{-1}\left| \frac{\tau_2}{\tau_1} \right| \text{sgn}(\tau_2) \\
    \theta_\Pi = \text{tan}^{-1}\left| \frac{\Pi_2}{\Pi_1} \right| \text{sgn}(\Pi_2).
    \end{split}
\end{equation}
Note that $\kappa=0.41$ is set for all TRIS simulations. The expressions for these terms are the following,
\begin{equation}
    \text{ln} Re_{\ast} = \kappa \left| \left( \frac{\pi^2}{4} + 1 \right) \left< \widetilde{u_i} \right>_0 - \frac{\pi^2}{4} \left< \widetilde{u_i} \right>_1 \right| - \kappa B + \frac{\pi^2}{8} + 1,
    \label{eq:lnRe}
\end{equation}
\begin{equation}
    e_{i,\ast} = \frac{ \left( \frac{\pi^2}{4} + 1 \right) \left< \widetilde{u_i} \right>_0 - \frac{\pi^2}{4} \left< \widetilde{u_i} \right>_1}{\left| \left( \frac{\pi^2}{4} + 1 \right) \left< \widetilde{u_i} \right>_0 - \frac{\pi^2}{4} \left< \widetilde{u_i} \right>_1 \right|},
    \label{eq:delta_star}
\end{equation}
\begin{equation}
    \Pi = \left|\frac{\pi^2}{4} \kappa ( \left< \widetilde{u_i} \right>_1 -  \left< \widetilde{u_i} \right>_0) - \frac{\pi^2}{8} e_{i,\ast}  \right|,
    \label{eq:Pi}
\end{equation}
\begin{equation}
    e_{i,\Pi} = \frac{1}{\Pi} \left( \frac{\pi^2}{4} \kappa ( \left< \widetilde{u_i} \right>_1 -  \left< \widetilde{u_i} \right>_0) - \frac{\pi^2}{8} e_{i,\ast} \right).
\end{equation}
With this formulations, the local values of the shear stress and top velocity are
\begin{equation}
    \widetilde{\tau_i} = \left( \frac{Re_\ast}{Re_{\tau}} \right)^2 e_{i,\ast},
    \label{eq:tau_i}
\end{equation}
\begin{equation}
    U_{T,i} = \left[ \frac{1}{\kappa} \text{ln} Re_{\ast} + B  \right] e_{i,\ast} + \frac{2 \Pi}{\kappa} e_{i,\Pi},
    \label{top_velocity}
\end{equation}
respectively. The pressure is closed by assuming a linear pressure profile that leads to the following relationship between the pressure boundary difference and pressure integral moment difference,
\begin{equation}
    \widetilde{p}_\text{top} - \widetilde{p}_\text{bot} = 6 \left[ \langle \widetilde{p} \rangle_1 - \langle \widetilde{p} \rangle_0 \right].
    \label{eq:pressure_difference}
\end{equation}
For simplicity, the wall-parallel nonlinear terms, $\left< \widetilde{u_i u_j} \right>_0$ and $\left< \widetilde{u_i u_j} \right>_1$, are split into the resolved and unresolved components captured by the Coles profile,
\begin{equation}
    \langle \widetilde{u_i u_j} \rangle_0 = \langle U_i U_j \rangle_0 + \sigma_{0,ij},
	\hspace{0.2\linewidth} 
    \langle \widetilde{u_i u_j} \rangle_1 = \langle U_i U_j \rangle_1 + \sigma_{1,ij},
    \label{eq:velocity decomposition}
\end{equation}
where $\sigma_{0,ij}$ and $\sigma_{1,ij}$ are defined as,

\begin{equation}
    \sigma_{0,ij} = C_s \Delta^2 \sqrt{\langle S_{mn} \rangle_{0} \langle S_{mn} \rangle_0} \langle S_{ij} \rangle_{0},
    \hspace{0.1\linewidth} 
    \sigma_{1,ij} = C_s \Delta^2 \sqrt{\langle S_{mn} \rangle_{1} \langle S_{mn} \rangle_1} \langle S_{ij} \rangle_{1}.
\end{equation}
Here, $S_{ij}$ is the strain rate tensor in the wall-parallel directions, $\Delta$ is the grid spacing of the TRIS simulation, and $C_s=0.78$ is set for the eddy viscosity coefficient. The conditional (resolved) components are quasi-linearized for robustness, defined as,
\begin{equation}
    \langle U_i U_j \rangle_0
    = A_0 (e_{i,\ast} e_{j,\ast}) 
    + B_0 (e_{i,\ast} e_{j,\ast} + e_{i,\Pi} e_{j,\Pi}) 
    + D_0 (e_{i,\Pi} e_{j,\Pi}),
\end{equation}
\begin{equation}
    \langle U_i U_j \rangle_1
    = A_1 (e_{i,\ast} e_{j,\ast}) 
    + B_1 (e_{i,\ast} e_{j,\ast} + e_{i,\Pi} e_{j,\Pi}) 
    + D_1 (e_{i,\Pi} e_{j,\Pi}),
\end{equation}
where
\begin{equation}
    A_0 = \left(2 T_{\text{ref}}
    - \frac{2}{\kappa} \right) T 
    + \left( \frac{2}{\kappa^2} 
    - T_{\text{ref}}^2 \right),
\end{equation}
\begin{equation}
    B_0 = \frac{\Pi}{\kappa} T_{\text{ref}} 
    + \frac{\Pi_{\text{ref}}}{\kappa} (T - T_{\text{ref}}) 
    - \frac{\Pi}{\kappa^2}\left(1 
    - \frac{\text{Si}(\pi)}{\pi} \right),
\end{equation}
\begin{equation}
    D_0 = \frac{3 \Pi_{\text{ref}}}{\kappa^2} \left( \Pi - \frac{1}{2} \Pi_{\text{ref}} \right),
\end{equation}
\begin{equation}
    A_1 = \left( 2 T_{\text{ref}}
    - \frac{1}{\kappa} \right) T 
    + \left( \frac{1}{2 \kappa^2} 
    - T_{\text{ref}}^2 \right),
\end{equation}
\begin{equation}
    B_1 = \left(1+ \frac{4}{\pi^2} \right) \left(\frac{\Pi}{\kappa}T_{\text{ref}} \right) 
    + \frac{\Pi_{\text{ref}}}{\kappa}(T - T_{\text{ref}}) 
    - \frac{2\Pi}{\kappa^2} \left( \frac{1}{4} - \frac{2 + \text{Cin}(\pi)}{\pi^2} \right),
\end{equation}
\begin{equation}
    D_1 = \left(3+\frac{16}{\pi}\right)\frac{\Pi_{\text{ref}}}{\kappa^2}\left(\Pi - \frac{1}{2}\Pi_{\text{ref}} \right).
\end{equation}
From these equations, $T = \frac{1}{\kappa} \text{ln}Re_{\ast} +B$ and the trigonometric integrals \citep{abramowitz1964} are defined as,
\begin{equation}
    \text{Si}(\pi)=\int_0^\pi\frac{\text{sin}(x)}{x}dx,
\end{equation}
\begin{equation}
    \text{Cin}(\pi)=\int_0^\pi\frac{\text{cos}(x)-1}{x}dx.
\end{equation}
$T_\text{ref} = \frac{1}{\kappa} \text{ln}Re_{\tau} +B$ and $\Pi_\text{ref}$ are the values at the reference condition. Specifically, $\Pi_\text{ref}$ and $B$ are set parameters matching the mean values of the streamwise velocity moments from DNS ($\langle \widetilde{u} \rangle_{0,\text{ref}}$ and $\langle \widetilde{u} \rangle_{1,\text{ref}}$), values listed in table \ref{tab:stat_comparison},
\begin{equation}
    \Pi_\text{ref}=\frac{\pi^2}{4}\kappa \left( \overline{\langle \widetilde{u} \rangle}_{1,\text{ref}} - \overline{\langle \widetilde{u} \rangle}_{0,\text{ref}} \right)
    -\frac{\pi^2}{8},
\end{equation}
\begin{equation}
    B=\left(\frac{\pi^2}{4}+1\right)\overline{\langle \widetilde{u} \rangle}_{0,\text{ref}}
    -\frac{\pi^2}{4}\overline{\langle \widetilde{u} \rangle}_{1,\text{ref}}
    +\frac{1}{\kappa}\left[ \frac{\pi^2}{8} + 1 - \text{ln} Re_\tau \right].
\end{equation}
The $\langle \widetilde{u_i v} \rangle_0$ term is closed by splitting the resolved and unresolved components, as done in \S\ref{sec:AMI},
\begin{equation}
    \langle \widetilde{u_i v} \rangle_0 = \langle \widetilde{u_i} \rangle_0 \langle \widetilde{v} \rangle_0 + \langle \widetilde{u''_i v''} \rangle_0,
\end{equation}
where $\langle \widetilde{u_i} \rangle_0=\langle U_i \rangle_0$ is known from \eqref{eq:Coles_mom0} and $\langle \widetilde{v} \rangle_0$ is known from the right equation in \eqref{eq:moments-of-mass}. The unresolved term, $\langle \widetilde{u''_i v''} \rangle_0$, is closed with an eddy viscosity approximation (with wall-normal scaling) superimposed with effects by the wake,

\begin{equation}
    -\langle \widetilde{u_i'' v''} \rangle_0
    \approx \int_0^1 \nu_{T,y} \frac{\partial U_i}{\partial y} dy
    + C_T \left( C_\Pi - 1 \right)\left[ \Pi e_{i,\Pi} - \Pi_\text{ref} e_{i,\ast} \right]
    \label{eq:unres_reyshear}
\end{equation}
where $\nu_{T,y}$ is an effective (dimensionless) wall-normal turbulent eddy viscosity, which is taken to be $\nu_{T,y} \approx C_T \kappa y$. Here, $C_T$ is a value that is soon to be defined and $C_\Pi$ is the tuning coefficient that controls the value of the resolved Reynolds shear stress. Using the skewed Coles profile, \eqref{eq:unres_reyshear} becomes the following,

\begin{equation}
    -2\langle \widetilde{u_i''v''} \rangle_0
    \approx C_T\left[1+(1-C_\Pi)\Pi_\text{ref}\right] e_{i,\ast}
    +C_T\left[C_\Pi \Pi\right] e_{i,\Pi}.
\end{equation}
At equilibrium ($\Pi=\Pi_\text{ref}$ and $e_{i,\ast}=e_{i,\Pi}$),
\begin{equation}
    -2\langle \widetilde{u_i''v''} \rangle_0
    =C_{uv}\delta_{i1}
    =C_T(1+\Pi_\text{ref})\delta_{i1},
\end{equation}
where $C_{uv}$ is a set parameter to control the amount of unresolved skin friction by turbulent enhancement. As a result, $C_T=C_{uv}/(1+\Pi_\text{ref})$. Therefore, the modeled unresolved Reynolds shear stress is defined as,

\begin{equation}
    -2\langle \widetilde{u''_i v''} \rangle_0 
    \approx C_{uv} \frac{\left[1+(1-C_{\Pi})\Pi_{\text{ref}}\right]e_{i,\ast} + C_{\Pi}\Pi e_{i,\Pi}}{1 + \Pi_{\text{ref}}}.
\end{equation}

\end{document}